\documentclass[a4paper,fleqn,usenatbib]{mnras}

\usepackage[T1]{fontenc}
\usepackage{ae,aecompl}

\usepackage{graphicx}	
\usepackage{amsmath}	
\usepackage{amssymb}	

\graphicspath{figs/}
\graphicspath{figs/caso2}

\title[High-energy radiation from HVCs]{High-energy radiation from collisions of high velocity clouds and the Galactic disk}

\author[del Valle, M\"{u}ller \& Romero]{
Maria V. del Valle,$^{1,2}$\thanks{E-mail: maria@iar-conicet.gov.ar (MVdV)}
A. L. M\"{u}ller$^{1,3,4}$
and G. E. Romero$^{1,5}$\\
$^{1}$Instituto Argentino de Radioastronom\'{\i}a (CONICET; CICPBA), C.C. No. 5, 1894 Villa Elisa, Argentina\\
$^{2}$Institute of Physics and Astronomy, University of Potsdam, 14476 Potsdam, Germany\\
$^{3}$Institut f\"{u}r Kernphysik (IKP), Karlsruhe Institut of Technology, Karlsruhe, Germany \\
$^{4}$Instituto de Tecnolog\'ias en Detecci\'on y Astropart\'iculas (CNEA, CONICET, UNSAM), Buenos Aires, Argentina\\
$^{5}$Facultad de Ciencias Astron\'omicas y Geof\'{\i}sicas, Universidad Nacional de La Plata, Paseo del Bosque s/n, 1900, La Plata, Argentina\\
}

\pubyear{2016}
\graphicspath{./figs/}

\begin{document}
\label{firstpage}
\pagerange{\pageref{firstpage}--\pageref{lastpage}}
\maketitle

\begin{abstract}
High-velocity clouds (HVCs) are interstellar clouds of atomic hydrogen that do not partake of the Galactic rotation and have velocities of a several hundred kilometers per second. A considerable number of these clouds are falling down towards the Galactic disk. HVCs form large and massive complexes, so their collisions with the disk must release a great amount of energy into the interstellar medium. The cloud-disk interaction produces two shocks, one propagates through the cloud and the other through the disk; the properties of these shocks depend mainly on the cloud velocity and the disk-cloud density ratio. In this work we study the conditions necessary for these shocks to accelerate particles by diffusive shock acceleration and the produced non-thermal radiation. We analyze particle acceleration in both the cloud and disk shocks. Solving a time-dependent 2-D transport equation for both relativistic electrons and protons we obtain particle distributions and  non-thermal spectral energy distributions. In a shocked cloud significant synchrotron radio  emission is produced along with soft gamma rays. In the case of acceleration in the shocked disk, the non-thermal radiation is stronger; the gamma rays, of leptonic origin, might be detectable with current instruments.  A large number of protons are injected into the Galactic interstellar medium, and locally exceed the cosmic-ray background. We conclude that under adequate conditions the contribution from HVC-disk collisions to the galactic population of relativistic particles and the associated extended  non-thermal radiation might be important.

\end{abstract}




\section{Introduction}

High-velocity clouds (HVCs) are a component of the neutral interstellar medium (ISM). These clouds are formed mainly by hydrogen and move with anomalous velocities (higher than the Galactic rotation velocity), with deviation velocities\footnote{$V_{\rm dev}$ is defined as the difference between the observed velocity of the gas and the maximum velocity expected from a simple model of differential galactic rotation. See \cite{wakker91}.} $V_{\rm dev} >90$~km~s$^{-1}$. There are at least three suggested origins for these clouds: 1) Material heated and transported to the Galactic halo by supernova explosions. Such material  cools and falls back to the Plane -- (this hypothesis is called \emph{Galactic fountain}). 2) Gas streams produced by tidal forces on nearby dwarf galaxies (e.g., the \emph{Magellanic Stream}). 3) Low-metallicity matter of intergalactic origin that falls onto the Galaxy. This latter low-metallicity component is thought to inject fresh material into the Galaxy for star formation and its existence is important to current models of Galactic evolution \citep[e.g.,][] {wakker13}. In any case, HVCs are essential for understanding the flows of energy and mass towards and within the Galaxy. In the past few decades their importance grew up considerably for two research fields: Galactic star formation and dark matter. This latter issue comes from the idea that at least some HVCs owe their existence to the presence of dark-matter halos \citep{blitz99,quilis01}.

A large fraction of HVCs have negative (approaching) velocities, so they will reach the Galactic plane at some time, colliding with the disk. The impact of these clouds with the gas in the disk should release a large amount of energy  into the ISM,  between $10^{47}-10^{52}$~erg. Such energetic events can trigger star formation episodes \citep{tenorio81}. The super massive HVC called the \emph{Smith  Cloud}, for example,  is thought to have collided with the Galactic disk  approximately 70~Myr ago. It should cross the plane again in about 27~Myr \citep{lockman08}. Numerical simulations of this collision suggest that, in order to have survived the impact, the Smith cloud should be embedded into a dark matter mini-halo \citep{nichols14}. \citet{galyardt16} studied the collision of this cloud and inferred the properties of the putative dark matter. High- and intermediate-velocity clouds can also  be used to trace cosmic rays (CRs) in the halo of the Milky Way \citep{tibaldo15}.
 
Several aspects of cloud-disk collisions as well as interactions of HVCs with their environment have been studied \cite[e.g., see Chapter 12 in][]{wakker13}.  The possibility of particle acceleration in these interactions, however, remains to be explored.  \citet{hedrick77} investigated the energy requirements for cosmic-ray acceleration in the inflow of HVCs material towards the Galactic plane. Collisions of HVCs with the Galactic disk were mentioned as potential CR sources for the first time in \citet{romero11}. In the current work we explore the particle acceleration that may take place in the shocks formed in the collisions and calculate the non-thermal radiative output. Preliminary results of this research were presented by \citet[][]{muller17}.

In the next section we briefly discuss the properties of the shocks created by the clouds when they impact on the disk and analyze the efficiency of  diffusive shock acceleration (DSA) in these shocks. In Section~\ref{sec:cloud} we present our model for the shock propagating through the cloud; we present and discuss the results of our calculations for this shock in Section\,\ref{sec:Results-cloud}. In the following Section\,\ref{sec:disk} we describe the model for the shock, moving through the disk. The results for the shocked disk scenario are presented in Section\,\ref{sec:Results-disk}. Finally, in Section~\ref{sec:concl} we give a short summary and offer our conclusions.

\section{Cloud-disk collisions}\label{cloud-disk}

High-velocity clouds have velocities in the range 100-500\,km\,s$^{-1}$ and typical densities between $0.1 $ and $1$ cm$^{-3}$. Even though the clouds are detected through the neutral H line, not all the gas is in neutral form. Ionized hydrogen is expected to constitute a large fraction of the material. Cloud radii vary greatly, from several pc to a few kpc, with a substantial number of clouds having radii greater than 50\,pc. HVCs are thought to form  large complexes. Metallicity and distance estimates are obtained through spectroscopy; the distance being one of the most difficult parameters to be determined for these objects. Concerning the metallicity, most clouds present sub-solar abundances. A complete description of HVCs and their characteristics is given by \citet{woerden04} and \citet{wakker13}.

If we consider a one-dimensional plane-parallel collision between a cloud of velocity $V_{\rm c}$ and density ${\rho}_{\rm c}$ with the disk of the Galaxy in a region where the density is ${\rho}_{\rm d}$, the velocities of the shocks moving through the cloud and the disk are \citep[see,][]{tenorio81,lee96}:

\begin{equation}\label{eq:vsc}
V_{\rm sc} = - \frac{4}{3} \;\frac{1}{1 + \sqrt{\rho_{\rm c}/\rho_{\rm d}}}\,V_{\rm c},
\end{equation}

\begin{equation}\label{eq:vsd}
V_{\rm sd} =  \frac{4}{3} \;\frac{1}{1 + \sqrt{\rho_{\rm d}/\rho_{\rm c}}}\,V_{\rm c}.
\end{equation}

\noindent In getting these expressions we have adopted  an adiabatic index $\gamma_{\rm gas} = 5/3$. Once the cloud's velocity $V_{\rm c}$ is fixed, the shock velocities depend only on the relative cloud-disk density. For moderate density contrasts the shock velocities are of the order of 100\,km\,s$^{-1}$.

\subsection{Particle acceleration}\label{sec:dsa}

 First order Fermi mechanism\footnote{ First order Fermi mechanism is a process by which charged particles gain energy by successive shock crossings, being scattered at both sides of the shock by magnetic fluctuations \citep[e.g.,][]{1978MNRAS.182..147B}. This process is thought to be the main mechanism for the production of CRs in the Galaxy.} or diffusive shock acceleration is known to efficiently operate in very fast shocks, with velocities of the order of $10^{-2}$~$c$ or higher. Recently, observations and numerical experiments suggest that first order Fermi mechanism can also operate in slower shocks, with velocities $\sim$ $10^{-3}$~$c$ \citep[e.g.,][]{caprioli14,lee15,metzger15}. This seems to be supported by the detection of synchrotron radiation from young stellar objects \citep{2010Sci...330.1209C,2016ApJ...818...27R}. There are, however, some limitations on DSA mechanism in slow shocks that are discussed below.

Shocks can be \emph{radiative} or \emph{adiabatic} depending on the efficiency of the gas to lose energy through thermal radiation. In radiative shocks the shocked material rapidly cools down resulting in large compression factors. A very dense gas layer forms and the shock promptly slows down \citep[e.g.,][]{drake05}. The efficiency $\eta$ to accelerate particles depends strongly on the shock velocity $V_{\rm s}$, $\eta$ $\propto (V_{\rm s} / c)^2$, hence this efficiency decays very fast in these shocks (although DSA in slightly radiative shocks might still operate). Moreover, if the post-shock  density attains very high values, collisions and ionization losses might be catastrophic for particle acceleration (see below). An adiabatic shock, on the other hand, propagates with approximately constant velocity through large spatial scales and its particle acceleration efficiency remains more or less constant. Because of this we consider only adiabatic shocks in what follows.

In order to determine the nature of a shock we compare the characteristic time scale $t_{\rm char}$ of the physical processes involved with the time scale of thermal losses $t_{\rm rad}$. If $t_{\rm char}$ $\ll$ $t_{\rm rad}$ the shock is  adiabatic. The time scale $t_{\rm rad}$ can be calculated as: 
\begin{equation}\label{eq:rad-loss}
t_{\rm rad} = \frac{5}{3} \frac{P}  {\mathcal{L}},
\end{equation}
\noindent where $P$ is the post-shock gas pressure, $T$ is the post-shock temperature $T = 2\times 10^{-9}V_{\rm s}^{2}$~K (with $V_{\rm s}$ in cm\,s$^{-1}$),  and $\mathcal{L} = n^{2}\Lambda (T)$; $n$ is the number density and $\Lambda (T)$ is the cooling function that can be fitted as a power-law in $T$ \citep[e.g.,][]{myasnikov98}\footnote{We adopt solar abundances, even though some clouds might have sub-solar metallicities, in which case $t_{\rm rad}$ is a lower limit.}.

In the absence of significant magnetic field pressure shocks cannot accelerate particles if their Mach number $\mathcal{M} = {V_{\rm s}} / {C_{\rm s}}$  $\leq \sqrt{5}$ and  $\mathcal{M}\leq 6$ for fully relativistic particles \citep{vink2014}, where $C_{\rm s}$ is the sound speed. The shock velocity then should be higher than  $6\,C_{\rm s}$. Even in the case of slow shocks, this constraint is not a limitation in this study because in the regions we are dealing with the temperatures are relatively low and the sound speed is much smaller than $100$ km\,s$^{-1}$.

When the ambient gas is very dense  or the shock velocity is not high enough, ionization  and Coulomb  losses of low-energy particles can be catastrophic, halting the particle acceleration process. If effective, the acceleration has to be fast enough at supra-thermal energies to compete with the collisional losses. The losses will not suppress the acceleration at any energy if \citep{drury96}:

\begin{eqnarray}\label{eq:colli}
&&\left(\frac{V_{\rm s}}{ 10^{3}\, {\rm km}\,{\rm s}^{-1}}\right)^2\left( \frac{B}{1\,\mu{\rm G}}\right)  \left(\frac{n}{1\,\rm{cm}^{-3}}\right)^{-1} \gg \\
&& 10^{-6}{\rm Max}\left[ {\chi}_{\rm i}T_{4}^{-1/2}, (1 - {\chi}_{\rm i})\right].\nonumber
\end{eqnarray}

\noindent Here $n$ is the number density,  $B$ is the magnetic field, $T_{4}$ is the gas temperature in units of $10^4$\,K and  ${\chi}_{\rm i}$ the ionization fraction.

Another limitation for DSA in cold media is due to ion-neutral friction. A cold medium is not fully ionized and hence a large number of neutral atoms and molecules exists in the gas, the ion-neutral friction damps the turbulent inhomogeneities with which particles scatter during the acceleration process. Depending on the degree of ionization these interactions can limit particle acceleration. Ion-neutral wave damping places no restriction on shock acceleration if the following condition is satisfied upstream \citep{drury96}:

\begin{equation}\label{eq:damp-cond}
\left(\frac{V_{\rm s}}{ 10^{3}\, {\rm km}\,{\rm s}^{-1}}\right)^3 \gg 8\times 10^{-3}\left( \frac{B}{1\,\mu{\rm G}}\right)^2 \left(\frac{n_{\rm n}}{1\,{\rm cm}^{-3}}\right) \left(\frac{n_{\rm i}}{1\,{\rm cm}^{-3}}\right)^{-2};
\end{equation}

\noindent here $n_{\rm n}$ and $n_{\rm i}$ are the neutral and ion number density respectively. When the latter condition is not fulfilled particles would accelerate but only until they reach a \emph{break momentum}. The maximum energy, in units of particle rest mass energy $mc^2$, is given approximately by \citep{malkov11}: 

\begin{equation}\label{eq:damp-energy}
{E_{\rm max-fric}} /\,{mc^2} \sim 10 \left( \frac{B}{1\,\mu{\rm G}}\right)^2 T_{4}^{-0.4} \left(\frac{n_{\rm n}}{1\,{\rm cm}^{-3}}\right)^{-1} \left(\frac{n_{\rm i}}{1\,{\rm cm}^{-3}}\right)^{-1/2}.
\end{equation}

\subsection{Models}

We consider two cases of shocks moving through the cloud (reverse shocks), models \emph{CI} and \emph{CII}, and one case in which a strong shock, induced by a HVC collision,  propagates through the disk (a forward shock), model \emph{D}. We are interested in studying strong adiabatic shocks, hence we focus on clouds with high velocities $\sim$ 500~km\,s$^{-1}$, as $V_{\rm s}$ $\propto$ $V_{\rm c}$ (see Eqs.~\ref{eq:vsc} and~\ref{eq:vsd}). Table~\ref{table:one} shows the main parameters of the models.

\begin{table}
\caption{Models parameters.}
\begin{center}
\begin{tabular}{lllll}
\hline
Model & $n_{\rm c}$ & $n_{\rm d}$ &  $V_{\rm s}$  & $\mathcal{M}$ \\
 &  [cm$^{-3}$] & [cm$^{-3}$]  &   [km~s$^{-1}$] &  \\
\hline
\emph{CI} & $0.1$ & $1.0$ & 500 & 43 \\
\emph{CII} & 0.5 & 0.1 & 200 & 16 \\
\emph{D} & 1.0 & 1.0 & 500  & 50 \\
\hline
\end{tabular}
\end{center}
\label{table:one}
\end{table}

In the following sections we describe the models and we analyze the properties of the shocks; first we deal with the shocks in the cloud (models \emph{CI} and \emph{CII}) and then we discuss the shock propagating through the disk (model \emph{D}).

\section{Model: Shocked HVC}\label{sec:cloud}

\begin{figure}
\begin{center}
\includegraphics[scale=0.7,trim=0cm 0cm 0cm 0cm, clip=true,angle=0]{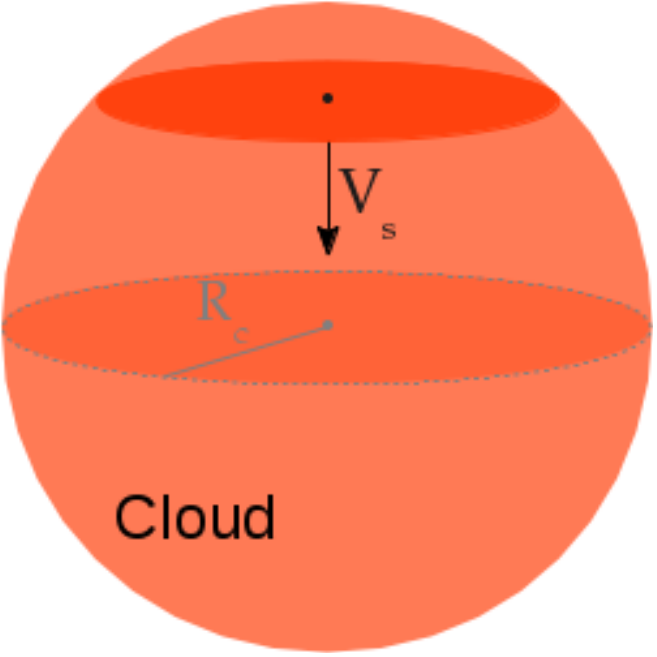}
\caption{Scheme of the physical scenario considered for the HVC (not to scale).}
\label{fig:cloud}
\end{center}
\end{figure}

We model the cloud as an homogeneous sphere of radius $R_{\rm c} = 10$~pc. We assume that protons and electrons are accelerated by the shock propagating through the cloud. The process occurs during the characteristic crossing time: $t_{\rm char} = 2 R_{\rm c}/ V_{\rm s}$. Figure~\ref{fig:cloud} illustrates this scenario.

\begin{figure}
\begin{center}
\includegraphics[scale=0.32,trim=0cm 0cm 0cm 0cm, clip=true,angle=270]{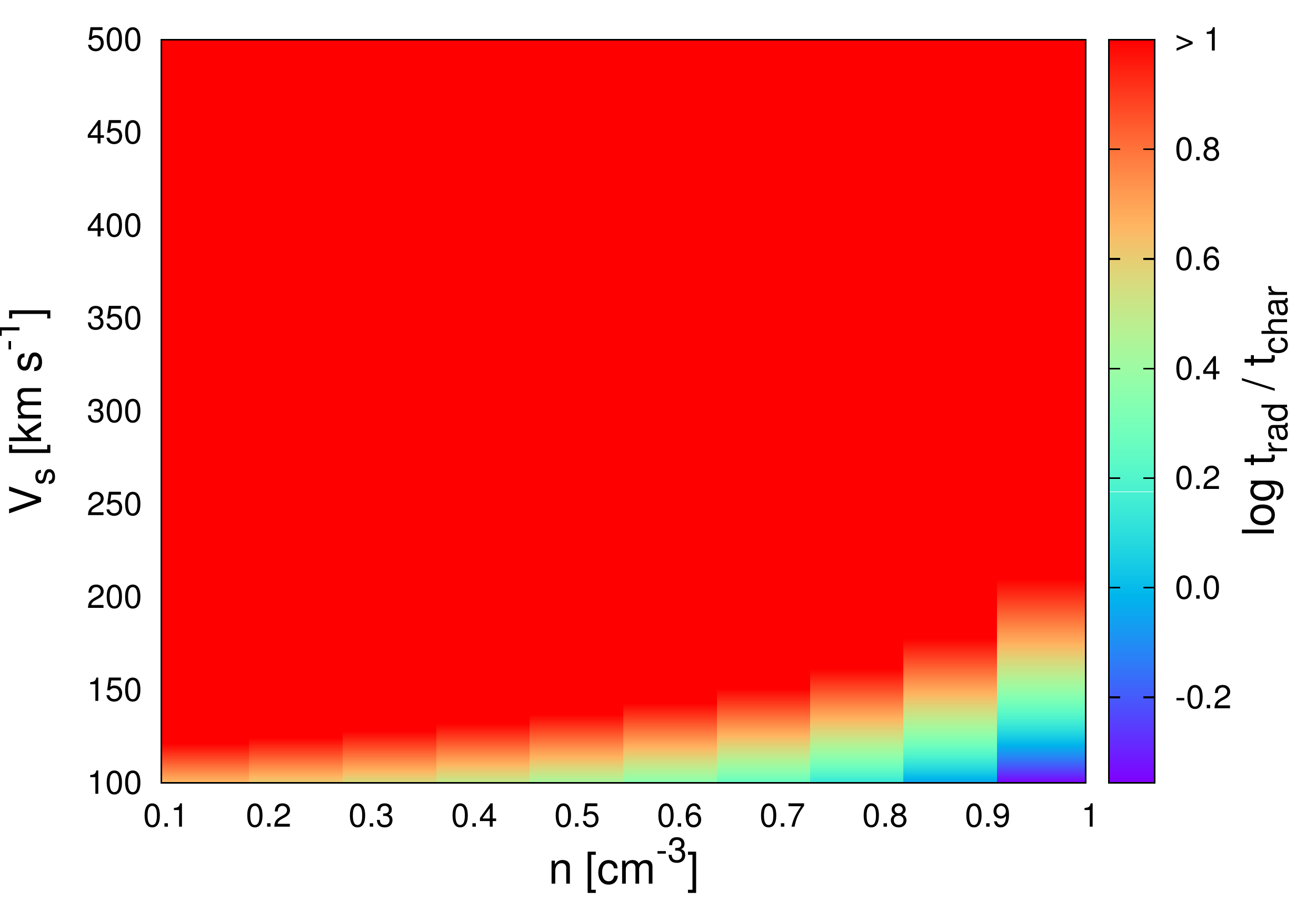}
\caption{$t_{\rm rad}/t_{\rm char}$ ratio as a function of density and shock velocity. The red region corresponds to $t_{\rm rad}/t_{\rm char}\geq 10$, i.e. adiabatic shocks.}
\label{fig:rad}
\end{center}
\end{figure}

In Fig.~\ref{fig:rad} we show the ratio $t_{\rm rad}/t_{\rm char}$  in logarithmic  scale as a function of density and shock velocity, see Eq.~(\ref{eq:rad-loss}); the regions where $t_{\rm rad}/t_{\rm char} \geq 10$ are shown in red. These are the regions where the radiative losses are highly inefficient. All three models lie in the adiabatic zone, so all shocks under consideration are adiabatic. The collision of the cloud with the disk also creates a forward shock (propagating through the disk), but this shock is radiative with the parameters adopted in \emph{CI} and \emph{CII}.  We then ignore the forward shocks in these models. 
 
Magnetic fields can play an important role in cloud dynamics and collisions. There is only one detection so far of magnetic fields in this type of objects that establishes a lower limit to the field on the line-of-sight of $B_{\rm lim} = 8$~$\mu$G \citep{hill13}. We assume $B = 10$~$\mu$G for both cloud models.

Since the clouds have relatively low densities, the condition given by expression ~(\ref{eq:colli}) is widely fulfilled. We adopt a typical value for the cloud temperature of $T_{\rm c} = 10^{4}$ K \citep{wakker13}. 
 
Calculating the maximum  energy that particles achieve is not simple because DSA is a non-linear process; we can obtain an estimate by computing the balance between the energy gain  and energy-loss rates. The relevant non-thermal radiative losses are synchrotron, relativistic Bremsstrahlung and inverse Compton (IC) scattering with the cosmic background radiation (CMB) at $T = $ 2.7\,K, for electrons, and $p-p$ inelastic collisions for protons. Particles might also escape from the acceleration region by diffusion; we adopt Bohm diffusion close to the shock. The values of $E_{\rm max-loss}$ obtained for electrons and protons are shown in Table~\ref{table:two}. Our quantitative estimates indicate that high energies are achievable (above 1 TeV).
 
The ionization degree of a cloud is not easy to estimate \citep[e.g.,][]{woerden04} and HVCs can vary from 99\% neutral to almost fully ionized. Sometimes a value $\chi_{\rm i} = 0.5$ is adopted. In other cases, however, there seems to be more ionized than neutral gas \citep{wakker08}. In the case of incomplete ionized clouds ion-neutral friction might impose the maximum particle energy.  In order to illustrate this last case we take here a representative  ionization degree of $\chi_{\rm i} = 0.5$. With this value condition ~(\ref{eq:damp-cond}) is not fulfilled and the maximum energies allowed are obtained using Eq.~(\ref{eq:damp-energy}). Since $E_{\rm max-fric}$ results lower than $E_{\rm max-loss}$ (see Table~\ref{table:two}),  ion-neutral friction halts the acceleration at the highest energies. We therefore adopt $E_{\rm max} = E_{\rm max-fric}$ as the more realistic estimate. 

\begin{table*}
\caption{Estimates of parameters in models \emph{CI} and \emph{CII}.}
\begin{center}
\begin{tabular}{lcccccc}
\hline
Model & \multicolumn{2}{c}{$E_{\rm max-loss}$ [GeV]} & \multicolumn{2}{c}{$E_{\rm max-fric}$ [GeV]} &  $L_{\rm par}$ [erg~s$^{-1}$] & $t_{\rm inj}$ [Myr]\\
 & e & p & e & p &  & \\
\hline
\emph{CI} & $5 \times 10^{3}$ & $4 \times 10^{5}$ & $10^{2}$ & $10^{5}$ & $2\times 10^{36}$ & $4\times10^{-2}$\\
\emph{CII}& $10^{3}$ & $10^{5}$ & $10^{1}$ & $10^{4}$ & $7\times 10^{35}$ & $10^{-1}$\\
\hline
\end{tabular}
\end{center}
\label{table:two}
\end{table*}

\subsection{Relativistic particle transport and emission}

The transport of relativistic protons and electrons is supposed to occur in the test-particle approximation. The spectral energy distribution $N_p$ of the particles obeys the equation: 
\begin{eqnarray}\label{eq:transcloud}
 \frac{\partial N_p}{\partial t}
= & D(E)\left[\frac{1}{R^2} \frac{\partial}{\partial R}
  \left( R^2 \frac{\partial N_p}{\partial R} \right) +  
\frac{1}{R^2{\sin}{\theta}} \frac{\partial }{\partial \theta} 
  \left( \sin \theta \frac{\partial \,N_p}{\partial \theta} \right)\right] \nonumber\\ 
  & - \frac{\partial}{\partial E} \left(P(R,\theta,E)\,N_p  \right)  
+ Q_p(R,\theta,E,t),
\label{ch7:esfer}
\end{eqnarray}

\noindent where $D(E)$ is the diffusion coefficient of the particles\footnote{We assume that the diffusion coefficient depends only on the particle energy.}, $P(\vec{r},E) \equiv -({\rm d}E/{\rm d}t)$ is the total radiative energy loss rate, and $Q_p(\vec{r},E,t)$ is the injection function. Given the geometry of the problem  we use a spherical coordinate system $(R,\theta,\phi)$, with its origin at the cloud center. The particle density function, $N_p$, depends spatially only on  $R$ and $\theta$, i.e. $N_p$ $\equiv$ $N_p(R,\theta,E,t)$. 

Relativistic particles, accelerated at the shock, are injected in a surface  $S_{\rm inj}= 2\pi r^2$, that is moving with velocity $V_{\rm inj} \equiv V_{\rm s}$ (see Fig.\,\ref{fig:cloud}). We consider that the particles have a power-law distribution in energy of index $\alpha = 2$, as expected from DSA in strong non-relativistic shocks. The injection function is normalized according to the power available in relativistic particles $L_{\rm par}$.   

The total kinetic power of the shock is estimated as $L_{\rm kin} = \frac{1}{2} \rho_{\rm c} V_{\rm s}^{2}\Omega_{\rm c}/t_{\rm char}$, where  $\Omega_{\rm c}$ is the volume of the cloud. The shocks in models \emph{CI} and \emph{CII} have Mach numbers $\mathcal{M}$ $\geq 10$, well above the limit  of 6 (see Table~\ref{table:two}). Numerical experiments show that non-relativistic shocks with such values of $\mathcal{M}$ can transform $10\%$ or more of their kinetic power into relativistic particles through DSA \citep[see,][]{caprioli14}, which is in agreement with other estimates \citep[e.g.,][]{1990ApJ...352..376E}. Here we take  $L_{\rm par} = 0.1 L_{\rm kin}$, with $L_{\rm par}$ equally divided between electrons and protons. 

Beyond some spatial scale the particle spatial diffusion changes from the Bohm regime in the acceleration region to a faster one, i.e. the diffusion coefficient increases. The acceleration process occurs within a region of linear size $l_{\rm acc} \sim D_{\rm Bohm}/V_{\rm s}$. For the maximum energies considered here $l_{\rm acc} \le 1$~pc.  Since we are modeling a cloud of 10~pc, we are considering phenomena occurring on scales 10 times larger. Consequently we expect the transition to a faster diffuse regime to occur within the cloud. We adopt a diffusion coefficient
\begin{equation}
  D(E) = 10^{26} \left(\frac{E}{10\,{\rm GeV}}\right)^{0.5}\,{\rm cm}^{2}\,{\rm s}^{-1},
\end{equation}  
\noindent similar to that of the ISM \citep[e.g.,][]{berezinskii90} but slightly smaller, because the magnetic field in the HVCs is greater than in the  ISM.

We solve Eq.~(\ref{eq:transcloud}) in a discrete grid  
$(E, R, \theta) \in [ 1 \; {\rm MeV}, \; 100 \; {\rm TeV} ] \times [ 0, \; 10 \; {\rm pc} ] \times [ 0, \; \pi ]$, using the finite-volumes method. The energy grid is logarithmically spaced, whereas the radial and polar grids are uniformly spaced. We use a  grid resolution $(L,M,K) = (64,32,32)$.  We integrate during $t_{\rm inj} \equiv t_{\rm char}$. For further description of the code see Appendix\,\ref{appendix}.

The particle distributions are interpolated into a 3D spatial grid. We calculate the non-thermal radiation produced by the particles as they diffuse through the cloud. We neglect the increase of the magnetic field or density due to compression by the shock. This latter simplification produces an underestimation of at most a factor of 4. The results and their descriptions are presented in the next section.

\section{Results: Shocked HVC}\label{sec:Results-cloud}
\graphicspath{./figs/figs/caso2/} 

\begin{figure*}
\begin{center}
\includegraphics[scale=1., clip=true,angle=0]{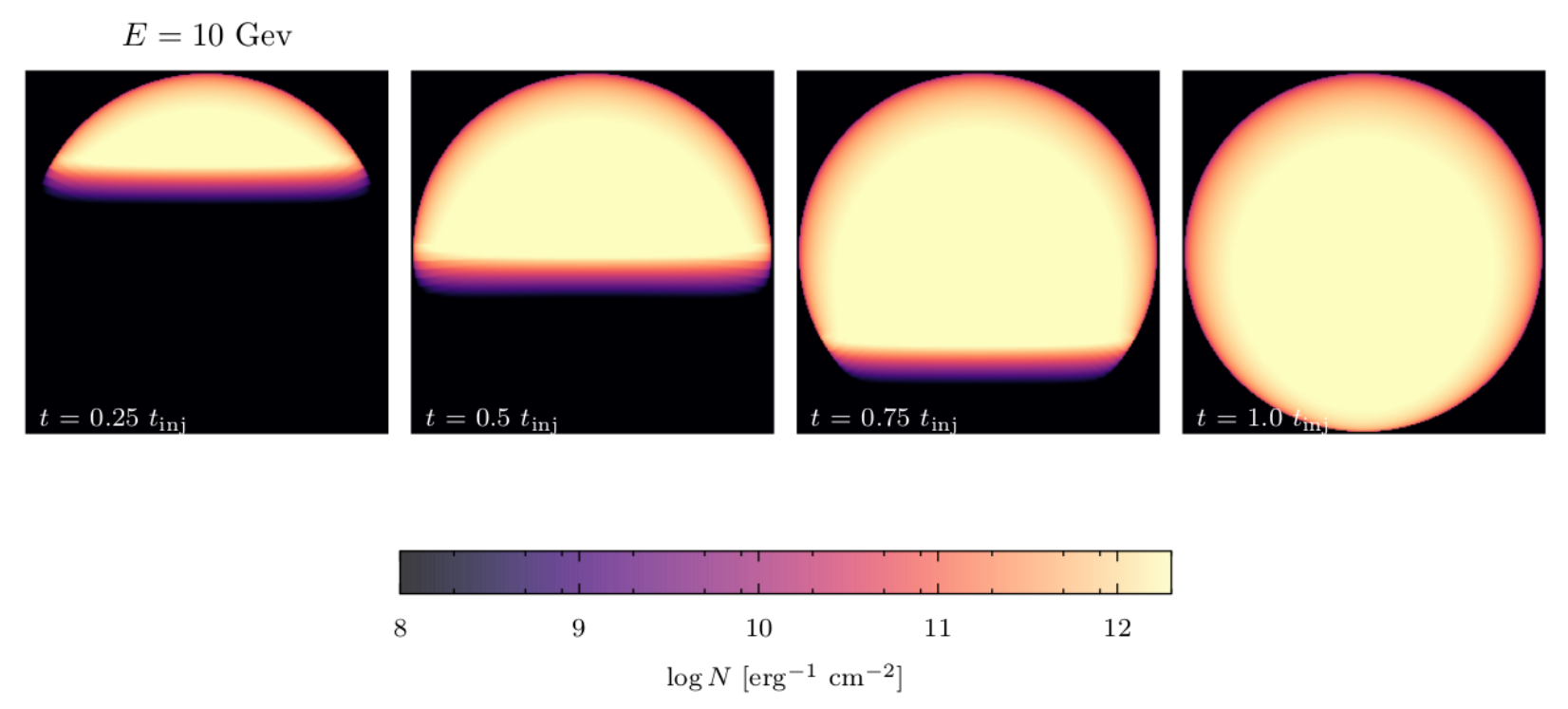}
\caption{Evolution of the number of protons with $E = 10$~GeV. These maps show the projection of the number of particles in the spherical cloud onto an arbitrary $x,y$-plane. Time evolves from left to right.}
\label{fig:Np-map}
\end{center}
\end{figure*}

The solution of the particle density distribution for protons of fixed energy $E = 10$~GeV is presented in Fig.~\ref{fig:Np-map}. This figure shows the projected $3D$ proton distributions in a $2D$ map, as a function of the injection time. The injected protons suffer energy losses due to $p-p$ inelastic collisions. The particle diffusion is not very fast at this energy, but its effects can be appreciated, especially in the last snapshots, in the shocked borders of the cloud. On these borders the number of particles decreases.

\begin{figure*}
\begin{center}
\resizebox{.67\columnwidth}{!}{\includegraphics[scale=.8,trim=0cm 0cm 0cm 2.2cm, clip=true,angle=270]{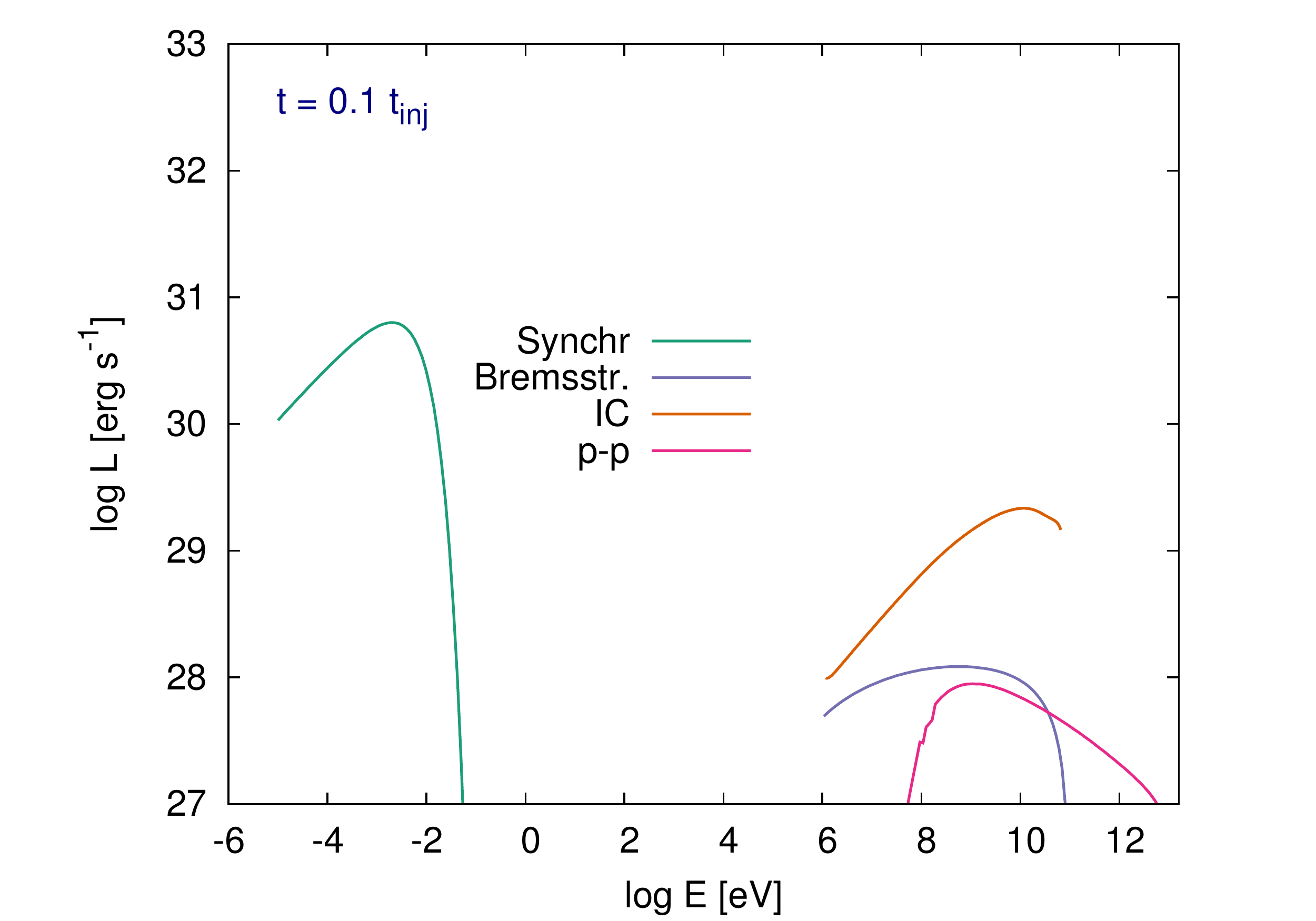}}
\resizebox{.67\columnwidth}{!}{\includegraphics[scale=.8, trim=0cm 0cm 0cm 2.2cm, clip=true,angle=270]{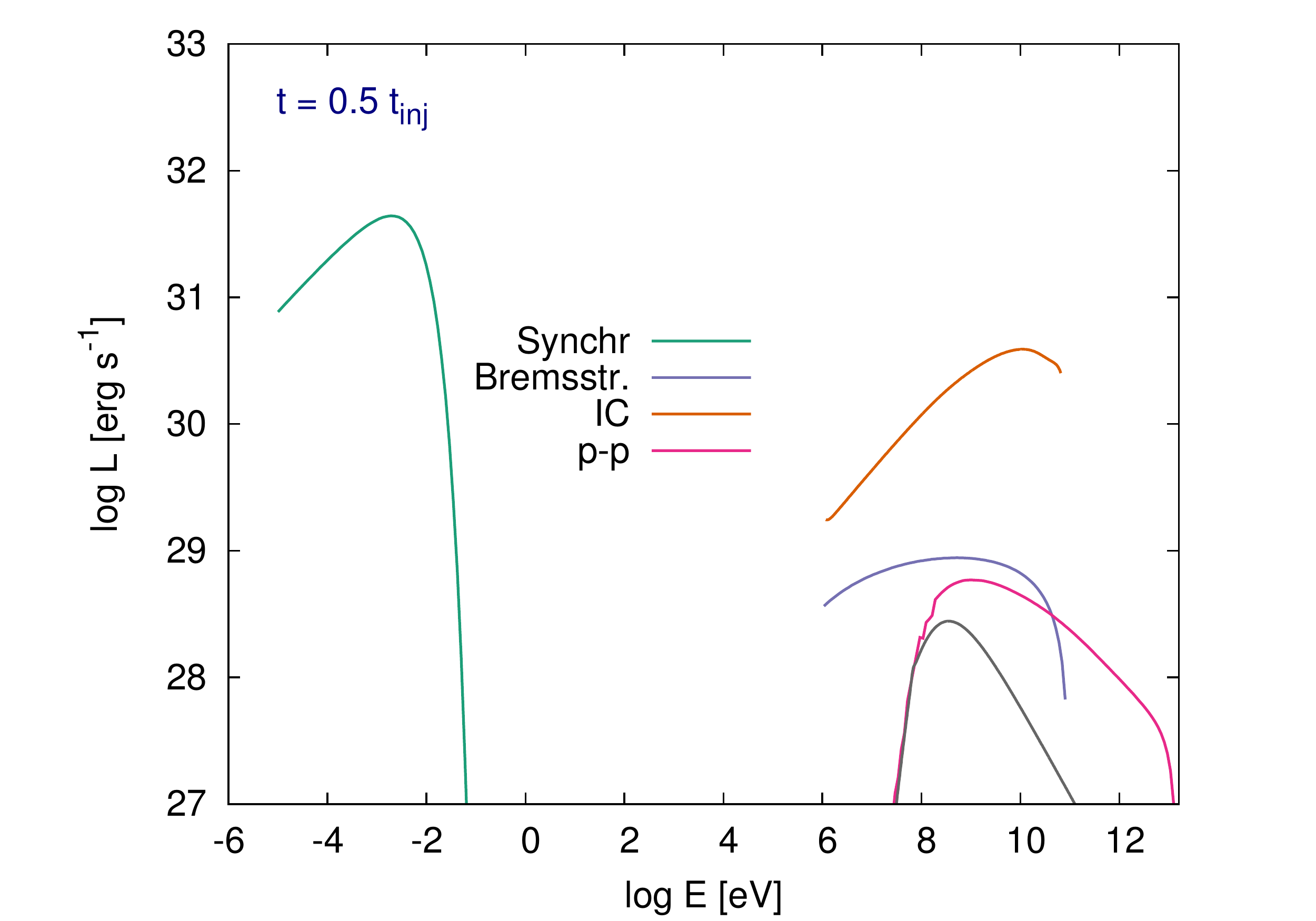}}
\resizebox{.67\columnwidth}{!}{\includegraphics[scale=.8, trim=0cm 0cm 0cm 2.2cm, clip=true,angle=270]{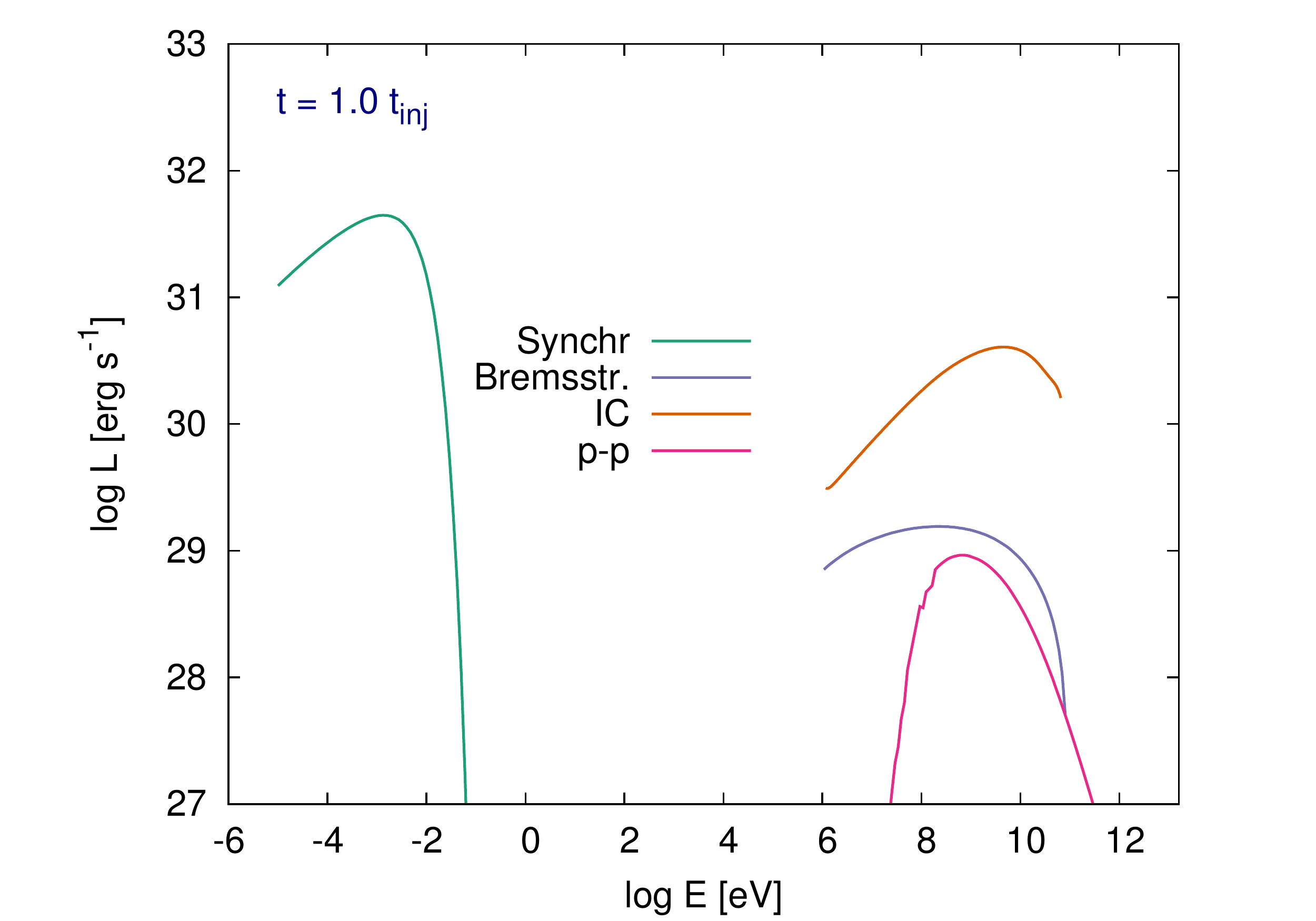}}\\
\caption{ Spectral energy distribution as a function of the injection time for Model \emph{CI}. The left panel shows the non-thermal SED for $t = 0.1\,t_{\rm inj}$, the middle panel shows the SED at  $t = 0.5\,t_{\rm inj}$ and the right panel corresponds to the final integration time $t = 1.0\,t_{\rm inj}$.}
\label{fig:sed-caso2}
\end{center}
\end{figure*}

The spectral energy distributions (SEDs) as a function of time are shown in Fig.~\ref{fig:sed-caso2}, for Model \emph{CI}. The non-thermal radio emission peaks at $E_{\rm ph} \sim 1.5\times 10^{-3}$~eV ($\equiv 380$~GHz) with luminosities greater than $10^{31}$~erg~s$^{-1}$. Between radio and soft gamma rays no significant non-thermal radiation is generated. The emission by protons reaches a maximum value at   $E_{\rm ph} \sim 5\times 10^{8}$~eV. However, in this region of the SED the contribution from the relativistic electrons dominates, with a peak at  $E_{\rm ph} \sim 10^{10}$~eV and luminosities $\sim$ $5\times10^{30}$~erg~s$^{-1}$. The part of the SEDs due to  electrons keeps approximately the same spectral index; in the case of the $p-p$ emission, after $t = 0.5~t_{\rm inj}$, the spectrum is modified: it gets steeper because the protons of the highest energies (those producing the radiation)  have left the cloud due to diffusion. At later injection times no emission is expected at  $E_{\rm ph} > 1$~TeV. The SED greater luminosities are not reached at $t = 1~t_{\rm inj}$, but shortly earlier. 

\graphicspath{./figs/figs/caso3/}

\begin{figure*}
\begin{center}
\resizebox{.67\columnwidth}{!}{\includegraphics[scale=.8,trim=0cm 0cm 0cm 2.2cm, clip=true,angle=270]{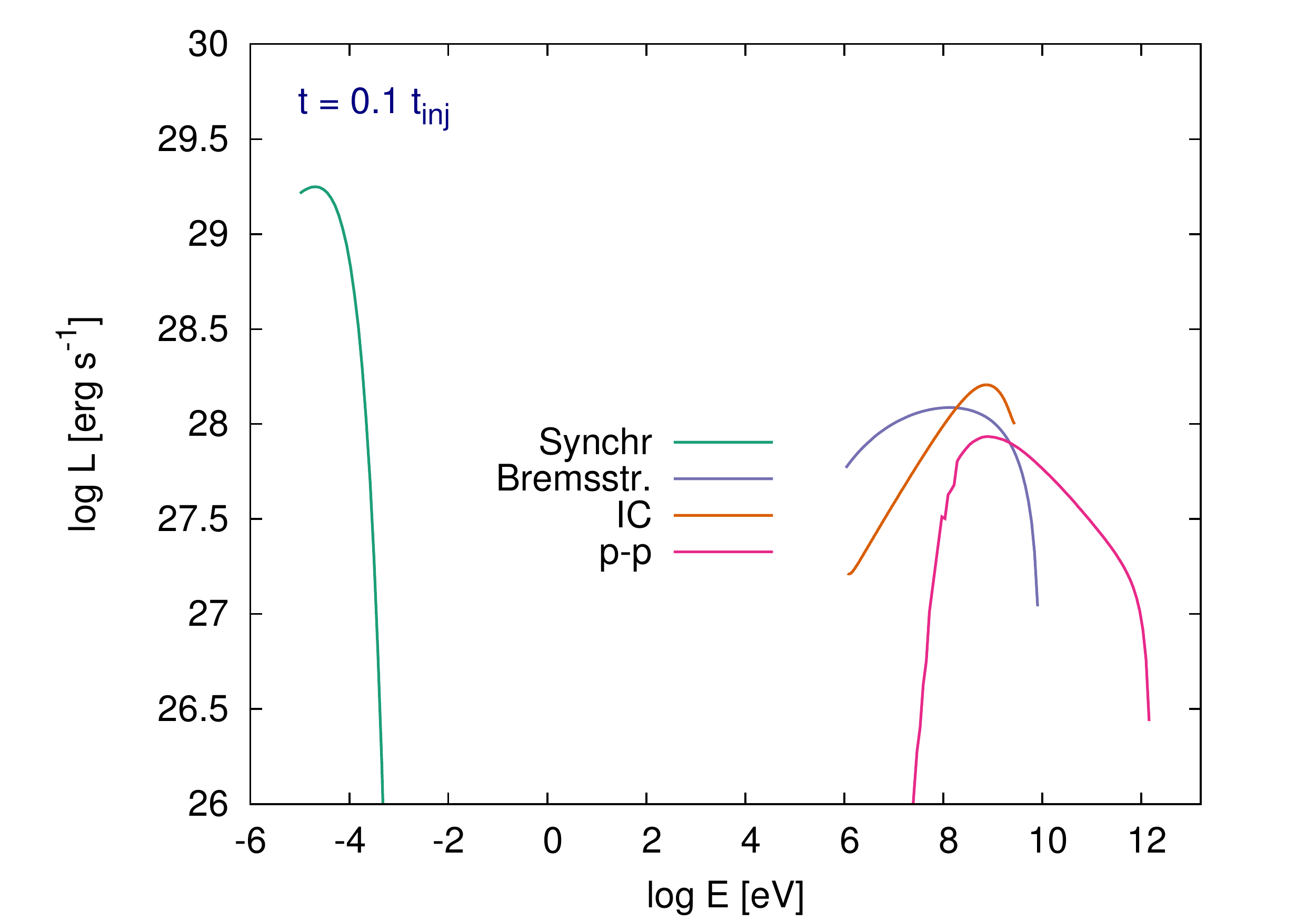}}
\resizebox{.67\columnwidth}{!}{\includegraphics[scale=.8, trim=0cm 0cm 0cm 2.2cm, clip=true,angle=270]{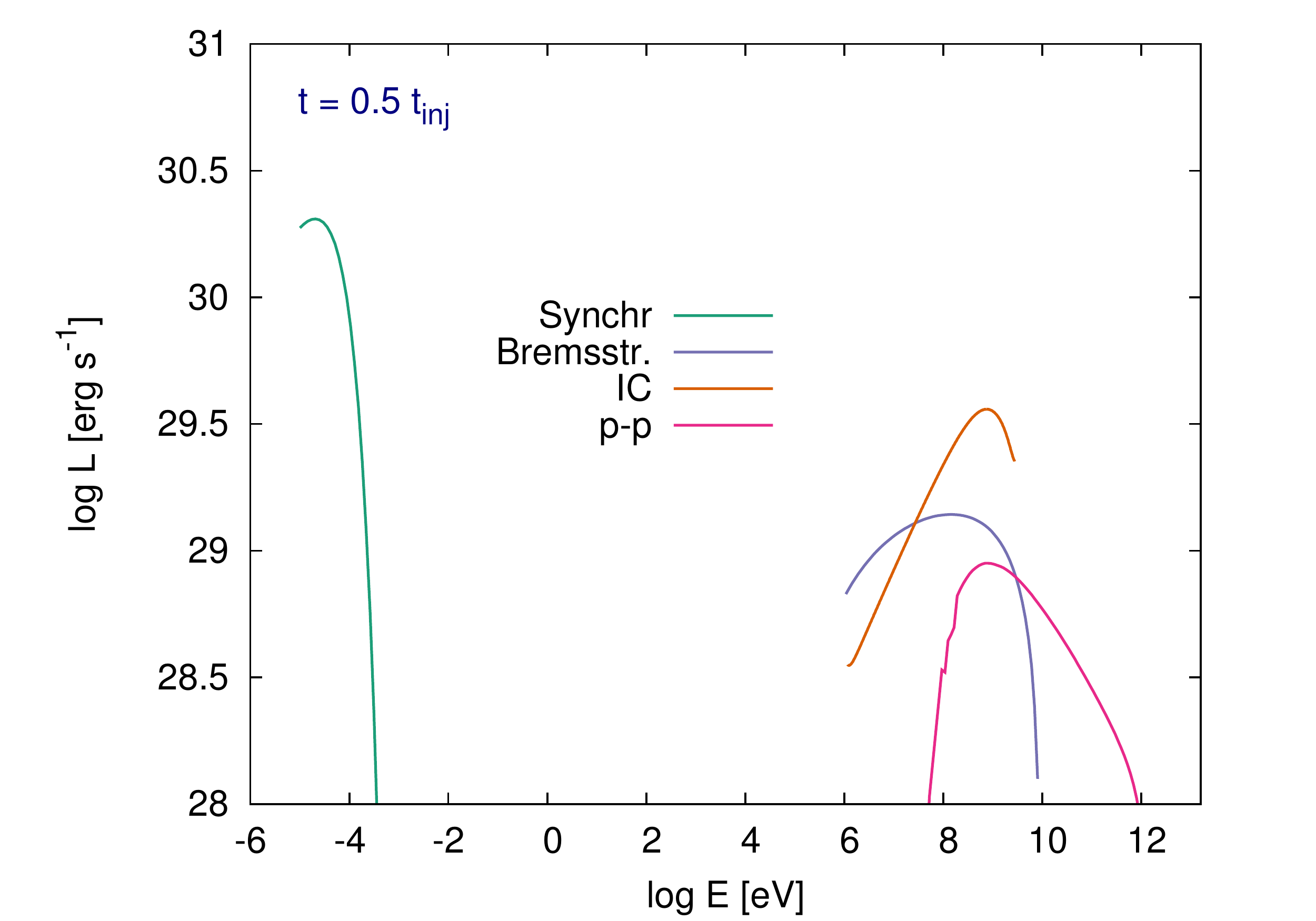}}
\resizebox{.67\columnwidth}{!}{\includegraphics[scale=.8, trim=0cm 0cm 0cm 2.2cm, clip=true,angle=270]{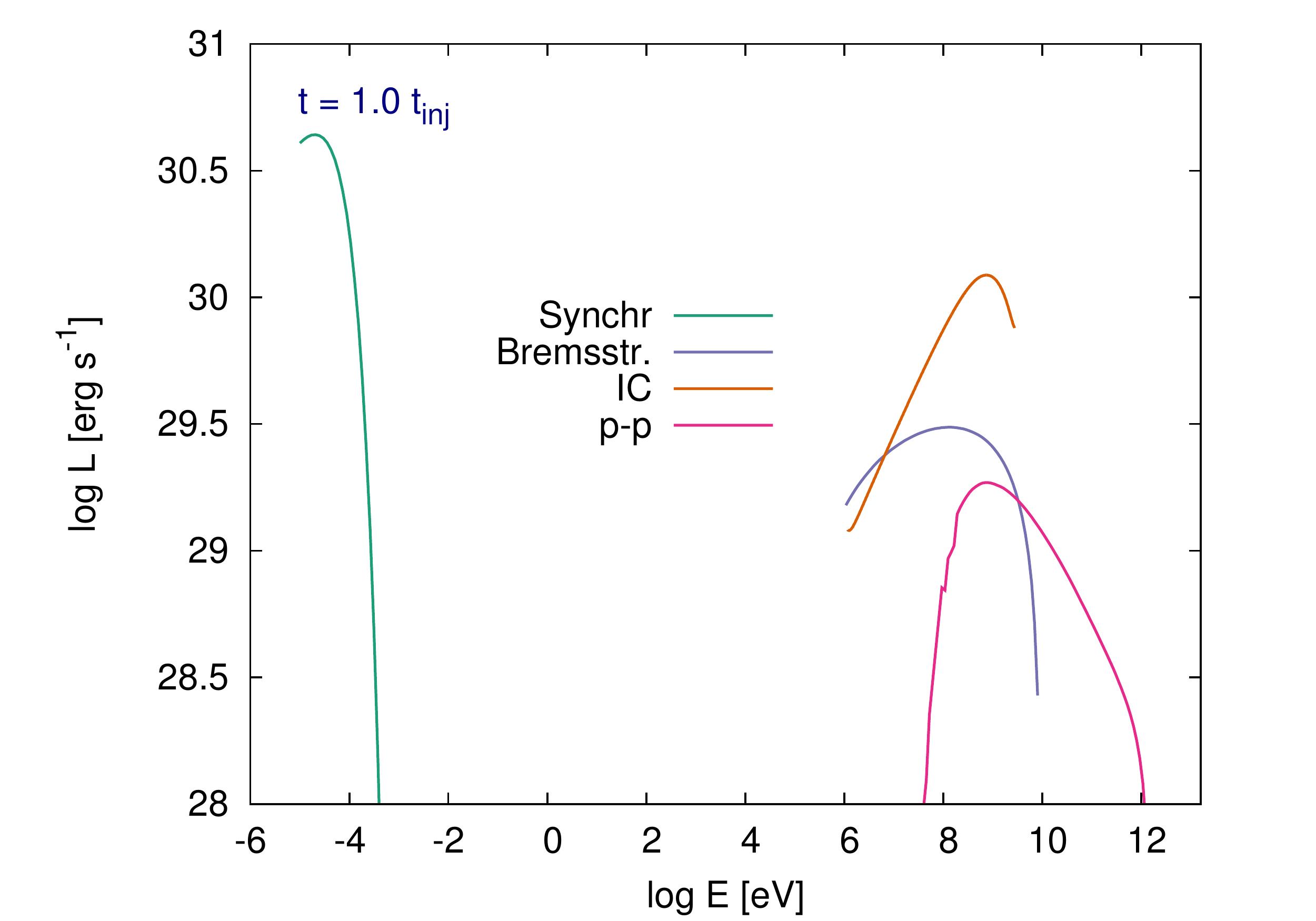}}\\
\caption{ Spectral energy distribution as a function of the injection time for Model \emph{CII}. The left panel shows the non-thermal SED for $t = 0.1\,t_{\rm inj}$, the middle panel shows the SED at  $t = 0.5\,t_{\rm inj}$ and the right panel corresponds to the final integration time $t = 1.0\,t_{\rm inj}$.}
\label{fig:sed-caso3}
\end{center}
\end{figure*}

The SEDs as a function of time for the Model  \emph{CII} are shown in Fig.~\ref{fig:sed-caso3}. In this case the luminosities are lower because of the smaller power in relativistic particles. These SEDs achieve the greater luminosities at the final integration time,  $t = 1~t_{\rm inj}$.  The non-thermal radio emission peaks at lower energies, $E_{\rm ph} \sim 3\times 10^{-5}$~eV ($\equiv 0.3$~GHz). The emission produced by protons reaches a maximum around the same energy as in the previous case. The gamma luminosities do not go beyond energies of 1~TeV and no steepening in the SEDs appears at high-energies. Also, IC radiation dominates the spectrum at soft gamma rays, with  a power in excess of $10^{30}$~erg~s$^{-1}$. At the highest energies,  $10^{10} < E_{\rm ph} < 10^{12}$~eV, the emission is greater than $10^{28}$~erg~s$^{-1}$.

\graphicspath{./figs/figs/caso2/} 

\begin{figure}
\begin{center}
\includegraphics[scale=.7,trim=0cm 0cm 0cm 0cm, clip=true,angle=0]{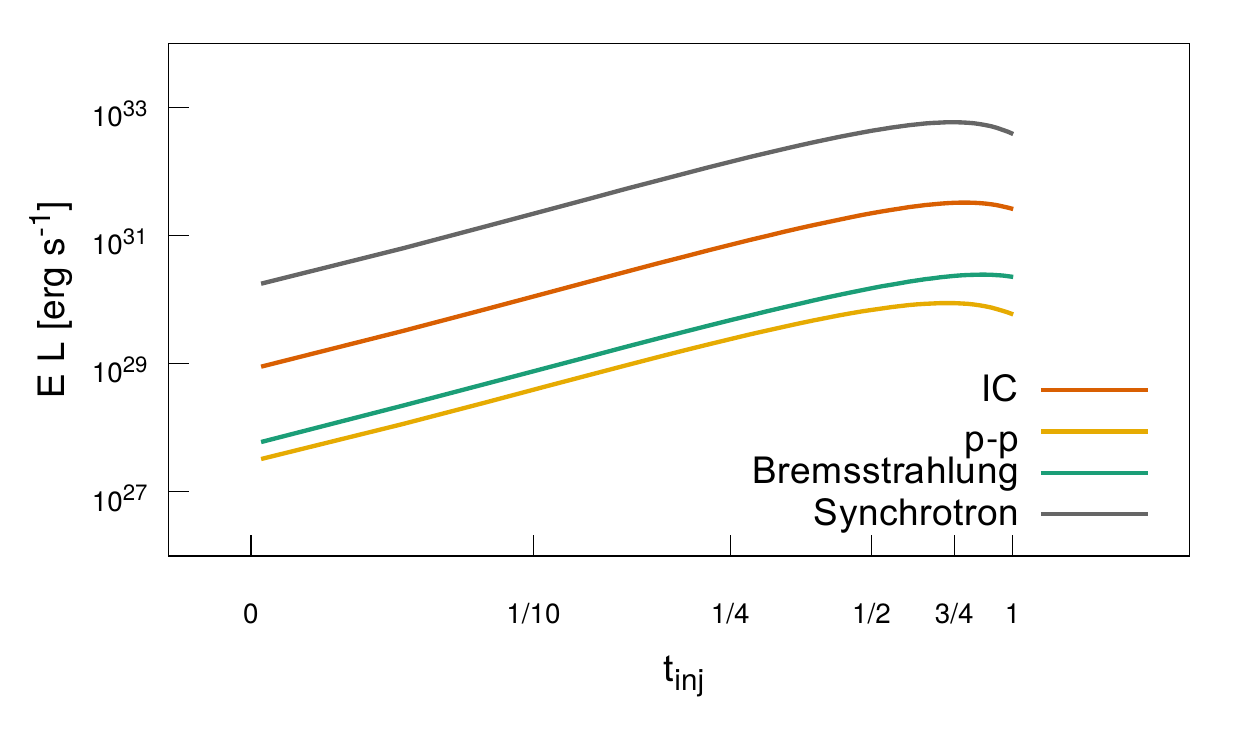}
\caption{Total luminosity as a function of time for the four main non-thermal radiative processes: synchrotron, IC scattering, relativistic Bremsstrahlung and $p-p$ interactions. This figure corresponds to Model \emph{CI}.}
\label{fig:total}
\end{center}
\end{figure}

In Fig.~\ref{fig:total} we show the time variation of the  total emitted power for the four main non-thermal radiative processes: synchrotron, relativistic Bremsstrahlung, IC scattering and $p-p$ interactions. This figure corresponds to Model \emph{CI}. The power grows slowly for the four mechanisms, varying almost two orders of magnitude between the initial time and the time when the maximum is reached. This maximum occurs almost at the same time for the leptonic processes, after the maximum of the $p-p$ emission. Clearly, synchrotron radiation is the most efficient non-thermal mechanism here, followed by IC. The maximum power emitted is $\sim 5.6\times10^{32}$erg~s$^{-1}$ for synchrotron. IC reaches $\sim 3\times10^{31}$erg~s$^{-1}$, followed by relativistic Bremsstrahlung that reaches $\sim 2.5\times10^{30}$erg~s$^{-1}$. Finally, we have $\sim 7\times10^{29}$erg~s$^{-1}$ for $p-p$. 

In the case of Model \emph{CII} (not shown here) the maximum luminosities of Bremsstrahlung and $p-p$ are greater, of $\sim 8.7\times10^{30}$erg~s$^{-1}$ and $\sim 2.9\times10^{30}$erg~s$^{-1}$, respectively. The total synchrotron and IC power are lower than in the case  \emph{CI}, being $\sim 3.4\times10^{31}$erg~s$^{-1}$ and $10^{31}$erg~s$^{-1}$, respectively.

\subsection{Discussion}\label{sec:discu}

HVCs form large complexes, with sizes 10 times, or more, the size of the individual clouds modeled here. A large cloud complex might fragment into smaller pieces during the collision process. In such a case the collective emission of all impacts can be one order of magnitude greater than the values we obtained for a single cloud. The non-thermal radio emission is the most significant radiative output produced in the collision. This emission peaks near 380\,GHz in the examples investigated here, far from  the observed radio emission of HVCs at 21~cm ($\sim$ 1.4\,GHz $\equiv$ $5.9 \times 10^{-6}$\,eV). The collective luminosity can be as high as  $\sim 5\times10^{33}$erg~s$^{-1}$, which is potentially detectable  considering distances of the order of the kpc.

The SED at soft gamma rays peaks near $E_{\rm ph} \sim 1$~GeV, close to the sensitivity peak of LAT instrument of the gamma satellite {\it Fermi}. For a source at $d \sim 1 $\,kpc, on the Galactic plane,  {\it Fermi} might detect sources at $E_{\rm ph} = 10^4$~MeV over $\sim 5\times 10^{31}$~erg~s$^{-1}$, for one calendar year all-sky survey\footnote{See \url{https://fermi.gsfc.nasa.gov/ssc/data/analysis/documentation/Cicerone/Cicerone_LAT_IRFs/LAT_sensitivity.html}}. In the collective case the luminosity might be detectable, and a longer integration time could result in a 5-$\sigma$ detection.

At higher energies, around  $E_{\rm ph}  \sim 1$~TeV, the system of Cherenkov telescopes MAGIC can detect a source with power over $\sim 2.3\times 10^{31}$~erg~s$^{-1}$  at 1~kpc \citep[see,][]{magic16}. The future array of Cherenkov telescopes CTA, at this same $E_{\rm ph}$, would have a sensitivity 1 order of magnitude higher; hence CTA  might be able to detect the gamma rays produced in the shocked cloud for a single cloud-disk collision.
 
The column densities of the HVCs are not too high. If the column density to the source (say a cloud in the Galactic plane at 1~kpc) is too high, the non-thermal emission produced by the Galactic CRs (electrons and protons) might be higher than the total luminosity of the shocked cloud, which will then be hidden by the background noise.

The proton cosmic-ray flux in the Galaxy is given by (e.g., \citealt{simpson83}):
\begin{equation}
\label{eq:CRs}
J_{\rm CR}^{\rm p}(E) = 2.2 \left(\frac{E}{\rm GeV}\right)^{-2.75}\,\,{\rm cm}^{-2}\,{\rm s}^{-1}\,{\rm sr}^{-1}\,{\rm GeV}^{-1}.
\end{equation}

\noindent For electrons we consider that $J_{\rm CR}^{\rm e}(E) = J_{\rm CR}^{\rm p}(E) / 100$, as indicated from observations \citep[e.g.,][]{berezinskii90}. Using these fluxes we computed the relativistic Bremsstrahlung and $p-p$ radiation expected from the background. For a column density of $ \sim 10^{19}$~cm$^{-2}$ \citep{1990ARA&A..28..215D} both emissions lie orders of magnitude lower than the contributions from the cloud, during practically all the integration time. In the case of a denser column density, $ \sim 10^{21}$~cm$^{-2}$, only the background $p-p$ emission is higher than that of the cloud; however this happens only at low energies,  between 10\,MeV and 10 GeV (see the grey-solid line in Fig.\,\ref{fig:sed-caso2}). Furthermore,  for $t \ge 0.5 t_{\rm inj}$, this noise gets completely under the $p-p$ cloud radiation.  

From plot\,\ref{fig:rad} it can be  seen that almost all the parameter space considered lies in the adiabatic region (red). We can have then  adiabatic shocks for a number of different parameters. Here we investigate the case of clouds with $V_{\rm c} = 500$\,km\,s$^{-1}$ as an extreme case, the majority of HVCs might have lower velocities. However, we analyze  qualitative what would be expected when  changing the density and the velocity of the cloud. 

In our model the cloud velocity and the density determine the shock velocity and the  power in relativistic particles. This last dependency is linear with the cloud density. A change in the power in relativistic particles is directly proportional to the non-thermal luminosity produced. The gamma emission from $p-p$ collisions is also linearly proportional to the density and will vary accordingly; however this is not the dominant non-thermal radiative process in these sources.   Given that the average number density of HVCs does not  change in many orders of magnitude, for a fixed shock velocity the number of relativistic particles is not expected to vary greatly from source to source.  

The velocity of the shock is a  more sensitive parameter of our model, and it is proportional to the cloud velocity. The shock velocity enters in the power in relativistic particles as $\propto V_{\rm s}^2$. For HVCs with velocities between 100 and 500\,km\,s$^{-1}$, and fixed density, the power in relativistic particles varies a factor of 25. This corresponds to a variation of a factor of 25 in the non-thermal emission. Also, the acceleration rate is $\propto V_{\rm s}^2$, and a variation of a factor of 25 is also expected in $E_{\rm max-loss}$. If the maximum energies are determined  by the losses and not by the ion-neutral damping (a highly ionized cloud) then the maximum energies the particles can reach vary in one order of magnitude.

We use here a ionization factor of $\chi_{\rm i} = 0.5$. For the parameters adopted in Models \emph{CI} and \emph{CII}, in the case of a fully or almost fully ionized cloud ion-neutral friction is not relevant, and the maximum energies are higher, given by $E_{\rm max-loss}$. In an almost complete neutral cloud the maximum energies particles attain would be slightly lower than those in Table\,\ref{table:two} (see dependencies in Eq.~(\ref{eq:damp-energy})).

We study here only adiabatic shocks because the DSA theory is well understood  in this regime. Radiative shocks are ubiquitous in our Galaxy; these shocks might also accelerate particles via DSA. However the radiation losses modify the shocks and studying the acceleration mechanism operating in such regime is very complex.   Assuming DSA in radiative shocks without a careful analysis would be too speculative. However, acceleration of particles and/or re acceleration of preexisting CRs in radiative shocks produced in  HVC-disk collision systems cannot be ruled out.

\section{Model: Shocked disk}\label{sec:disk}


\begin{figure}
\begin{center}
\includegraphics[scale=.8, trim=0.cm 0.cm 0.cm 0.cm, clip=true,angle=0]{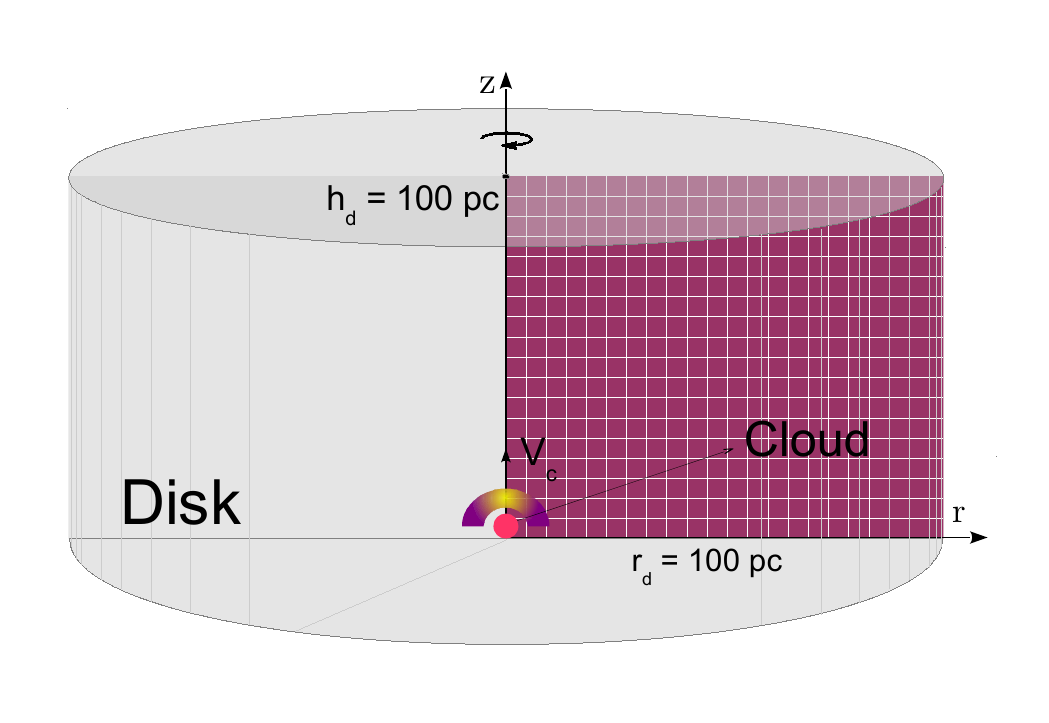}
\caption{Scheme of the physical scenario adopted for the Galactic disk (not to scale). The gridded purple surface indicates the computational plane $r, z$. The cloud-shock system is taken as a  punctual injector.}
\label{fig:disk}
\end{center}
\end{figure}

In order to model the effects of the collision of the disk,  we assume that the forward shock, which propagates through an homogeneous disk injects particles as a point source (see Fig.\ref{fig:disk}). The region of impact in the disk is modeled as a cylinder of radius $r_d = 100$~pc, height $h_d = r_d = 100$~pc and density $n_d = 0.1$\,cm$^{-3}$. The sound velocity in the warm ISM is of the order of 10\,km\,s$^{-1}$ \citep[e.g.,][]{1998ApJ...494L..19D}. The shock Mach number is then $\mathcal{M}$ $\sim$ 50 $\gg$ 6 (see Table\,\ref{table:one}). The characteristic crossing time is $t_{\rm char} = h_d / V_{\rm s} $, so $t_{\rm rad} \gg t_{\rm char}$ and the shock is adiabatic (see Sect.\ref{sec:dsa}). We adopt a magnetic field $B_{\rm d} = 4\, \mu$G for the compressed medium.

\begin{figure}
\begin{center}
\includegraphics[scale=.56, trim=0.cm 2.cm 0.cm 1.cm, clip=true,angle=0]{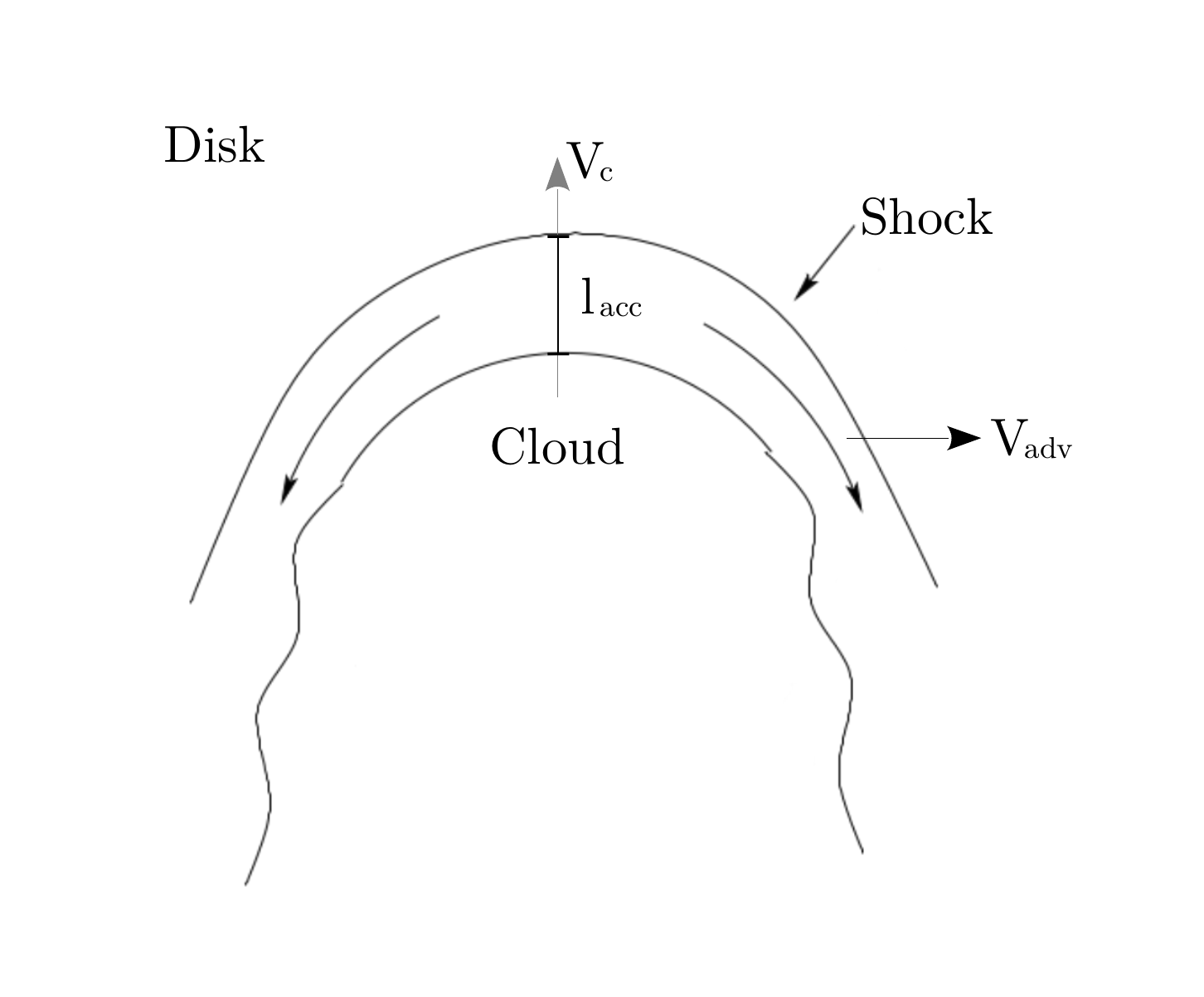}
\caption{Sketch of the acceleration scenario for the shock propagating through the disk,  model \emph{D} (not to scale).}
\label{fig:sketch}
\end{center}
\end{figure}

The forward shock injects protons and electrons along the characteristic time $t_{\rm char}$. A sketch of the acceleration scenario is shown in Fig.\,\ref{fig:sketch}\footnote{We only consider the details of the acceleration region for estimating the maximum energies. In the calculations of the transport of relativistic particles and non-thermal emission the source is treated as a punctual injector.}. As before, we estimate the particle maximum energies comparing the energy loss and gain rates.  The relevant non-thermal radiative losses  of electrons are due to synchrotron radiation, relativistic Bremsstrahlung, and IC scattering with the background radiation fields. The most relevant interstellar radiation fields are the CMB, the ambient infrared (IR) radiation field  (mainly produced by thermal emission from interstellar dust grains), and the ultraviolet (UV) contribution from the  integrated stellar radiation \citep[e.g.,][]{2013aism.book.....M}. Protons lose energy only through $p-p$ inelastic collisions. Particles, as in the previous cases, might escape the acceleration region, of size $l_{\rm acc}$, by diffusion and now they might also be drawn away from the acceleration zone, advected by the material that flows through the sides of the cloud with a velocity $V_{\rm adv}$ $\sim$ $V_{\rm s} / 4$ (Fig.\,\ref{fig:sketch}).

When a HVC approaches the disk it will find a composition of the different ISM phases that is mostly neutral, but with a strongly varying ($0-100$ \%) ionization fraction that is irregularly distributed on scales smaller than 100 pc. We consider here the limiting cases: a fully ionized disk and a 99\% neutral disk, i.e. $\chi_{\rm i}$ $\sim$ 0.01.

\begin{figure}
\begin{center}
\includegraphics[scale=.28, trim=0.8cm 1.2cm 0.8cm 1.2cm, clip=true,angle=0]{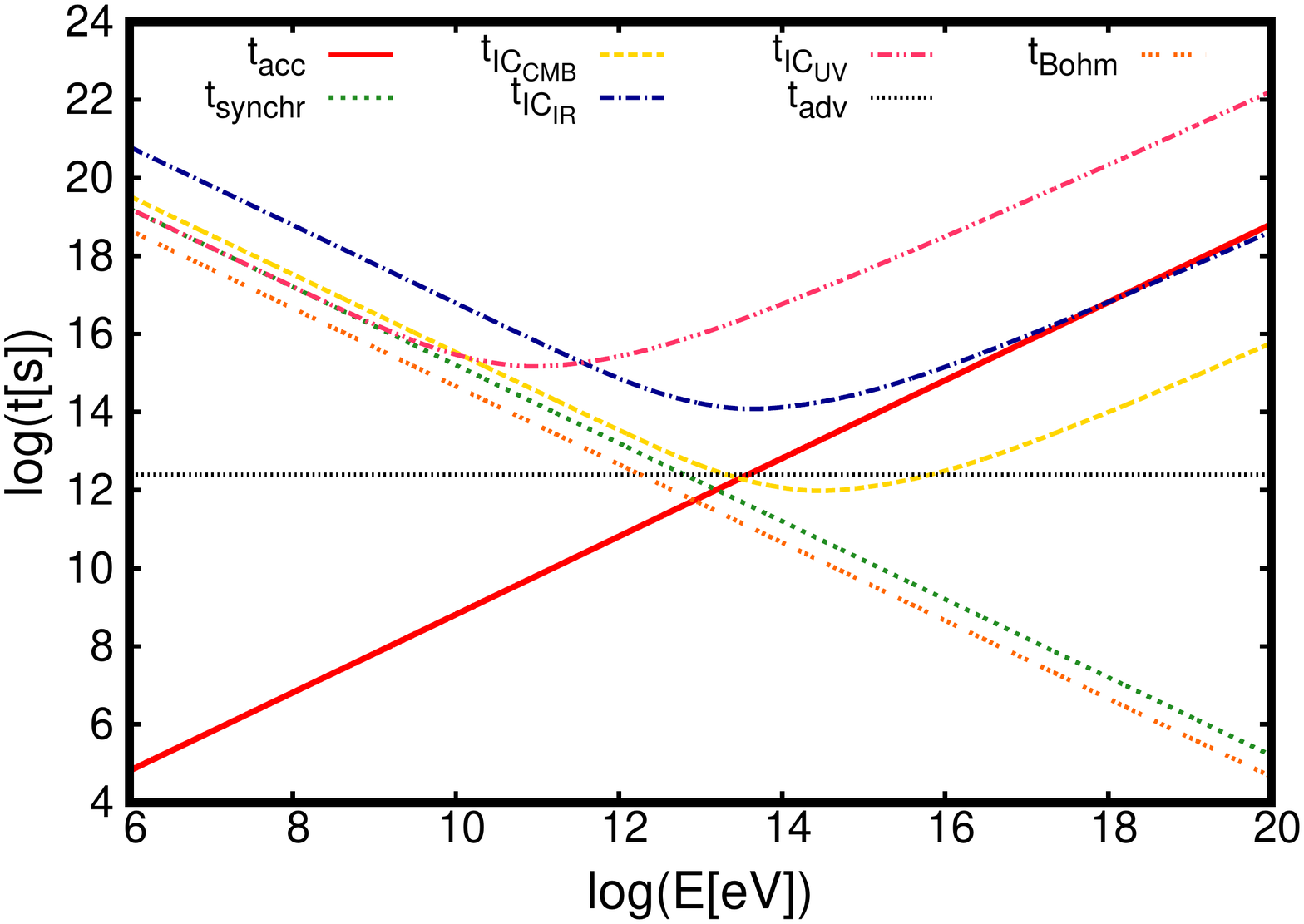}\\
\includegraphics[scale=.28, trim=0.8cm 1.2cm 0.8cm 1.2cm, clip=true,angle=0]{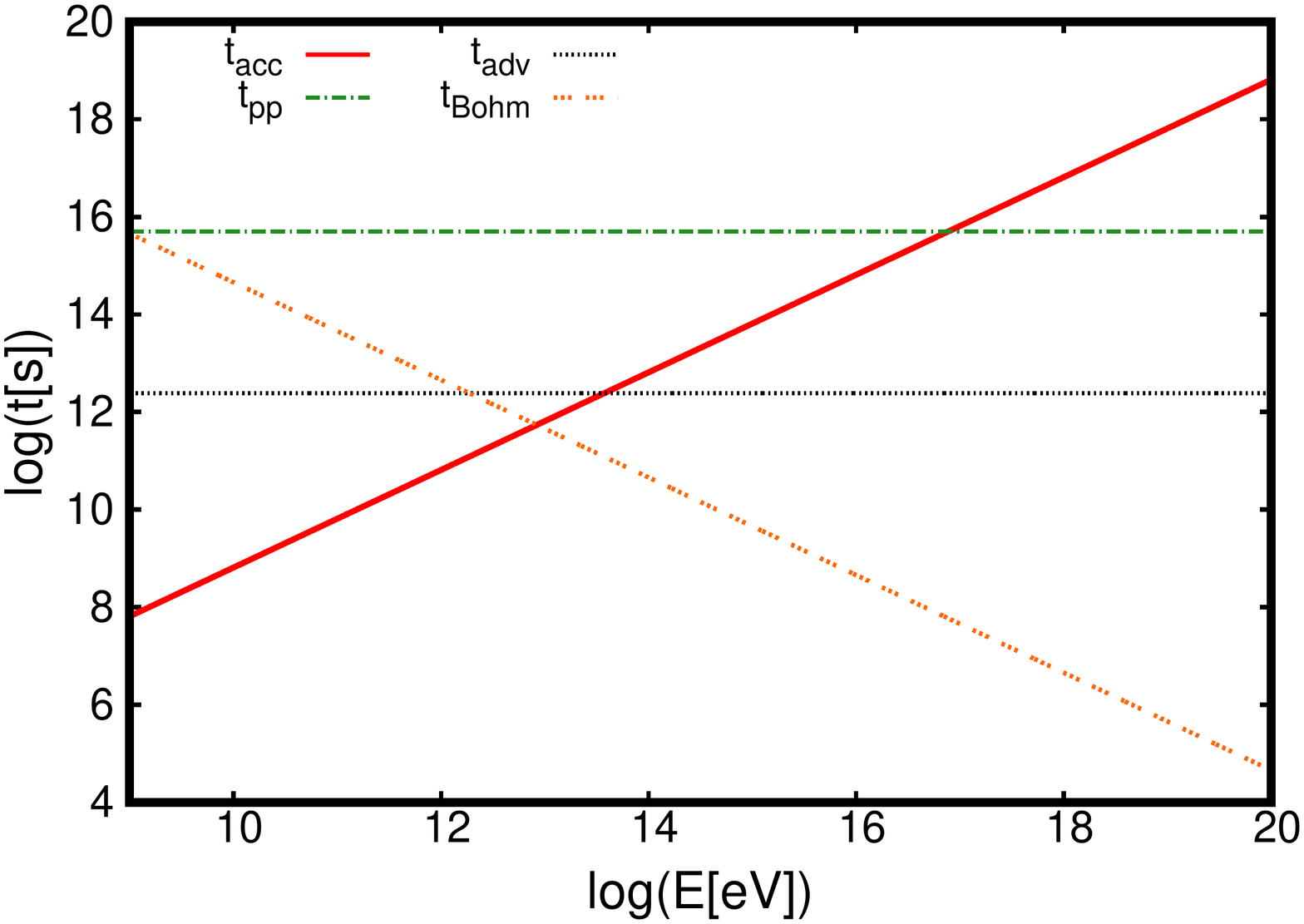}
\caption{Energy losses and acceleration time-scales of electrons (up) and protons (down), for model \emph{D}.}
\label{fig:losses}
\end{center}
\end{figure}

In Fig.\,\ref{fig:losses} we show the losses and acceleration time-scales of electrons (upper plot) and protons (bottom plot). For both species the shortest time-scale corresponds to diffusion. We then estimate the maximum energies  $E_{\rm max-loss}$ setting $t_{\rm acc} = t_{\rm Bohm}$; the  values  obtained for electrons and protons are shown in Table~\ref{table:three}. 
In the  case of a fully ionized disk ion-neutral damping does not occur but in the case with $\chi_{\rm i}$ $\sim$ 0.01 the maximum energies might be given by Eq.~(\ref{eq:damp-energy}). The  values  of $E_{\rm max-fric}$ for electrons and protons are also shown in Table~\ref{table:three}. We see that $E_{\rm max-fric}$ for protons is greater than $E_{\rm max-loss}$, and then the proton maximum energy for a neutral or fully ionized disk are identical. In the case of electrons, the maximum energy for $\chi_{\rm i}$ $\sim$ 0.01 is two orders of magnitude lower than in the case of a fully ionized disk.

\begin{table*}
\caption{Estimates of parameters in model \emph{D}.}
\begin{center}
\begin{tabular}{lcccccc}
\hline
\multicolumn{2}{c}{$E_{\rm max-loss}$ [GeV]} & \multicolumn{2}{c}{$E_{\rm max-fric}$ [GeV]} &  $L_{\rm par}$ [erg~s$^{-1}$] & $t_{\rm inj}$ [Myr]\\
 e & p & e & p &  & \\
\hline
$7 \times 10^{3}$ & $7 \times 10^{3}$ & $5\times 10^{1}$ & $5\times 10^{4}$ & $2\times 10^{36}$ & $2\times10^{-1}$\\

\hline
\end{tabular}
\end{center}
\label{table:three}
\end{table*}

\subsection{Relativistic particle transport and emission}\label{sec:part-disk}

A cylindrical coordinate system $(r,\;z)$ is the most natural one to describe particle transport in the disk. The $z$ axis is determined  by the direction of the cloud velocity. The transport equation  for relativistic protons and electrons, in cylindrical coordinates,  is: 

\begin{eqnarray}\label{eq:transdisc}
 \frac{\partial N_p}{\partial t}
= & D(E)\left[\frac{1}{r} \frac{\partial}{\partial r}
  \left( r \frac{\partial N_p}{\partial r} \right) +  
\frac{{\partial}^{2} N_p}{\partial z^{2}}\right] \nonumber\\ 
  & - \frac{\partial}{\partial E} \left(P(r,z,E,t)\,N_p  \right)  
+ Q_p(r,z,E,t),
\label{ch7:esfer}
\end{eqnarray}

\noindent where the terms represent the same as those of Eq.\,(\ref{eq:transcloud}). We solve the former equation in a $(r,\;z)$ plane (see Fig.\ref{fig:disk}) with $0\le z \le z_{\rm max} \equiv h_{\rm d}$ and $0\le r \le r_{\rm max} \equiv r_{\rm d}$. The energy  grid is the same as in the previous cases. We use a  grid resolution $(64,64,64)$. Again we integrate during $t_{\rm inj} \equiv t_{\rm char}$ (see Appendix\,\ref{appendix}).

The particles accelerated at the shock are injected in the disk, with a velocity $V_{\rm inj} \equiv V_{\rm s}$. We consider, as before, that the particles have a power-law distribution in energy of index $\alpha = 2$. Then the injection term is $Q_p(r,z,E,t) = Q_{\rm norm}\,E^{-\alpha}\delta({\vec{X}-\vec{X}_{\rm inj}})$, with $\vec{X}_{\rm inj}$ the shock position at time $t$, i.e. $\vec{X}_{\rm inj} = (0,V_{\rm inj}t )$.  $Q_{\rm norm}$  is the normalization factor, that depends on the power available in relativistic particles $L_{\rm par}$. We do not take into account any physical details of the shock+cloud system since we model the source as a point-like injector.  

A fraction of the  kinetic energy of the cloud is transferred to the forward shock in the collision. This fraction is quite large, so we approximate it as 1/2. Then, the power in the shock results $L_{\rm kin} = \frac{1}{4} \rho_{\rm c} V_{\rm s}^{2}\Omega_{\rm c}/t_{\rm char}$. As mentioned before $10\%$ or more of the shock power goes into relativistic particles in the DSA process. Consequently we adopt  $L_{\rm par} = 0.1 L_{\rm kin}$, with $L_{\rm par}$ equally divided between electrons and protons (see Table~\ref{table:three}). 

Beyond some spatial scale the diffusion coefficient becomes higher than in the Bohm approximation. Since we are interested in the large scale phenomena, we adopt a diffusion coefficient typical of the ISM:
\begin{equation}
  D(E) = 10^{27} \left(\frac{E}{10\,{\rm GeV}}\right)^{0.5}\,{\rm cm}^{2}\,{\rm s}^{-1}.
\end{equation}

With the time-dependent particle distributions obtained we calculate the non-thermal radiation produced by synchrotron, IC scattering, Bremsstrahlung, and by neutral pion decay due to $p-p$ inelastic collisions. 

We are interested in comparing the locally injected protons with that from the background CR population.  In order to obtain the background CR distribution consistently with the parameters adopted, we solve the transport equation in steady state with null injection function, with the initial condition $N^{\rm p}_{\rm CR} (t = 0) = 4\pi/c \,J_{\rm CR}^{\rm p}(E)$ (given in Eq.\,(\ref{eq:CRs})) and with boundary conditions matching the proton flux of the Galactic CR protons $J_{\rm CR}^{\rm p}$.

\section{Results: Shocked disk}\label{sec:Results-disk}
\graphicspath{./figs/figs/caso1/}

\begin{figure*}
\begin{center}
\includegraphics[scale=1., clip=true,angle=0]{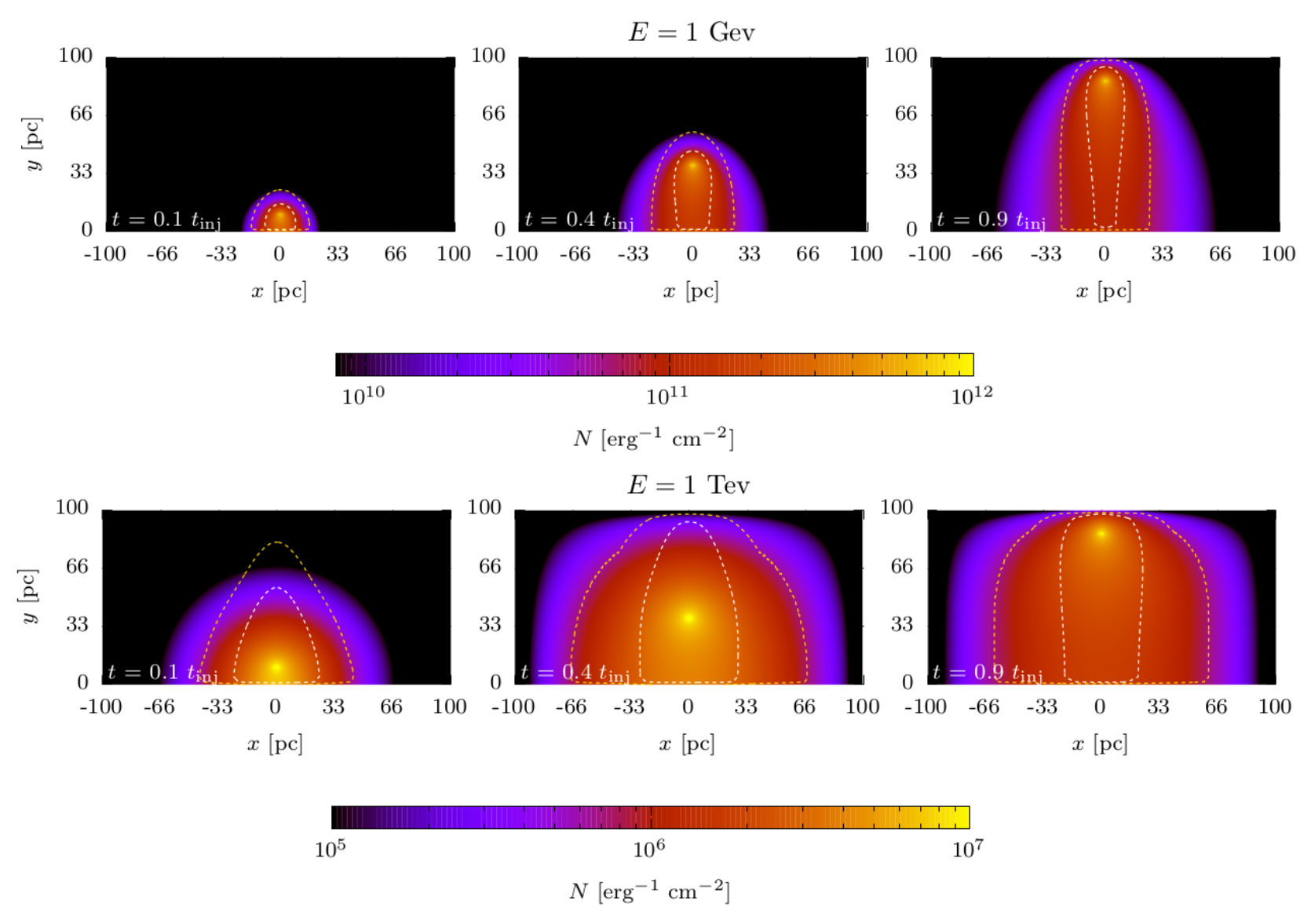}\\

\caption{Evolution of the number of protons with $E = 1$~GeV (top) and $E = 1$~TeV. These maps show the projection of the number of particles in the cylindrical disk onto a $x,y$-plane with $y$ in the direction of the cloud's motion. Time evolves from left to right.}
\label{fig:Np-map-disk}
\end{center}
\end{figure*}

Figure\,\ref{fig:Np-map-disk} shows maps of the evolution of protons with $E = 1$~GeV (top) and $E = 1$~TeV (bottom) as they are injected into the disk. In order to produce these maps we project the 3D space $(r,z,\phi)$, with $ 0 \le \phi \le 2\,\pi$ the azimuthal angle, into a $(x,y)$ Cartesian plane. As the particles are injected they diffuse through the disk; the more energetic particles diffuse faster. During the integration time, most of these particles do not reach the boundaries of the region, so they are concentrated around the axis in a zone of radius $\sim$ 20\,pc for the protons with $E = 1$~GeV, and of  $\sim$ 66\,pc for those with $E = 1$~TeV.

\begin{figure*}
\begin{center}
\resizebox{.67\columnwidth}{!}{\includegraphics[scale=.8,trim=0cm 0cm 0cm 2.2cm, clip=true,angle=270]{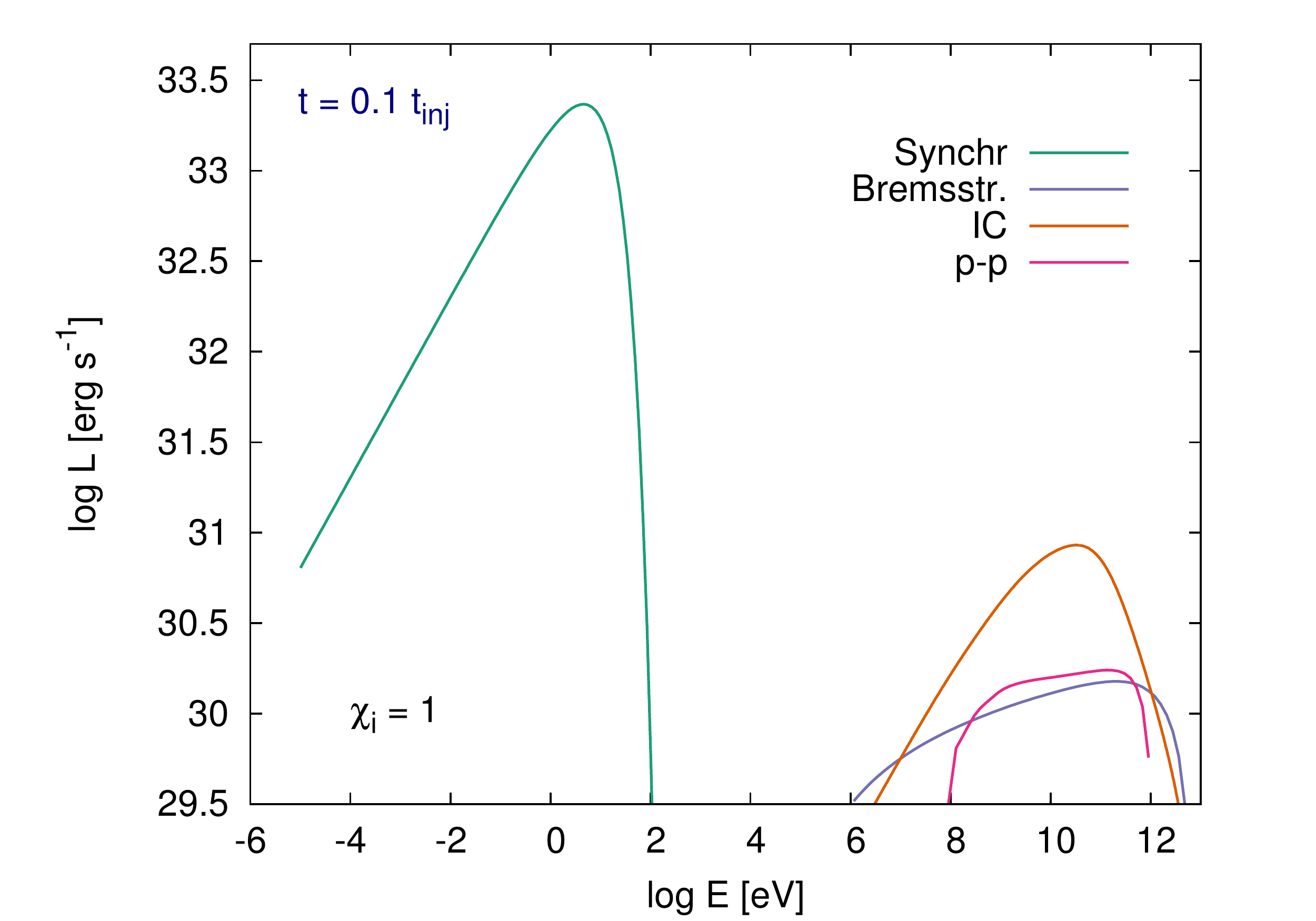}}
\resizebox{.67\columnwidth}{!}{\includegraphics[scale=.8, trim=0cm 0cm 0cm 2.2cm, clip=true,angle=270]{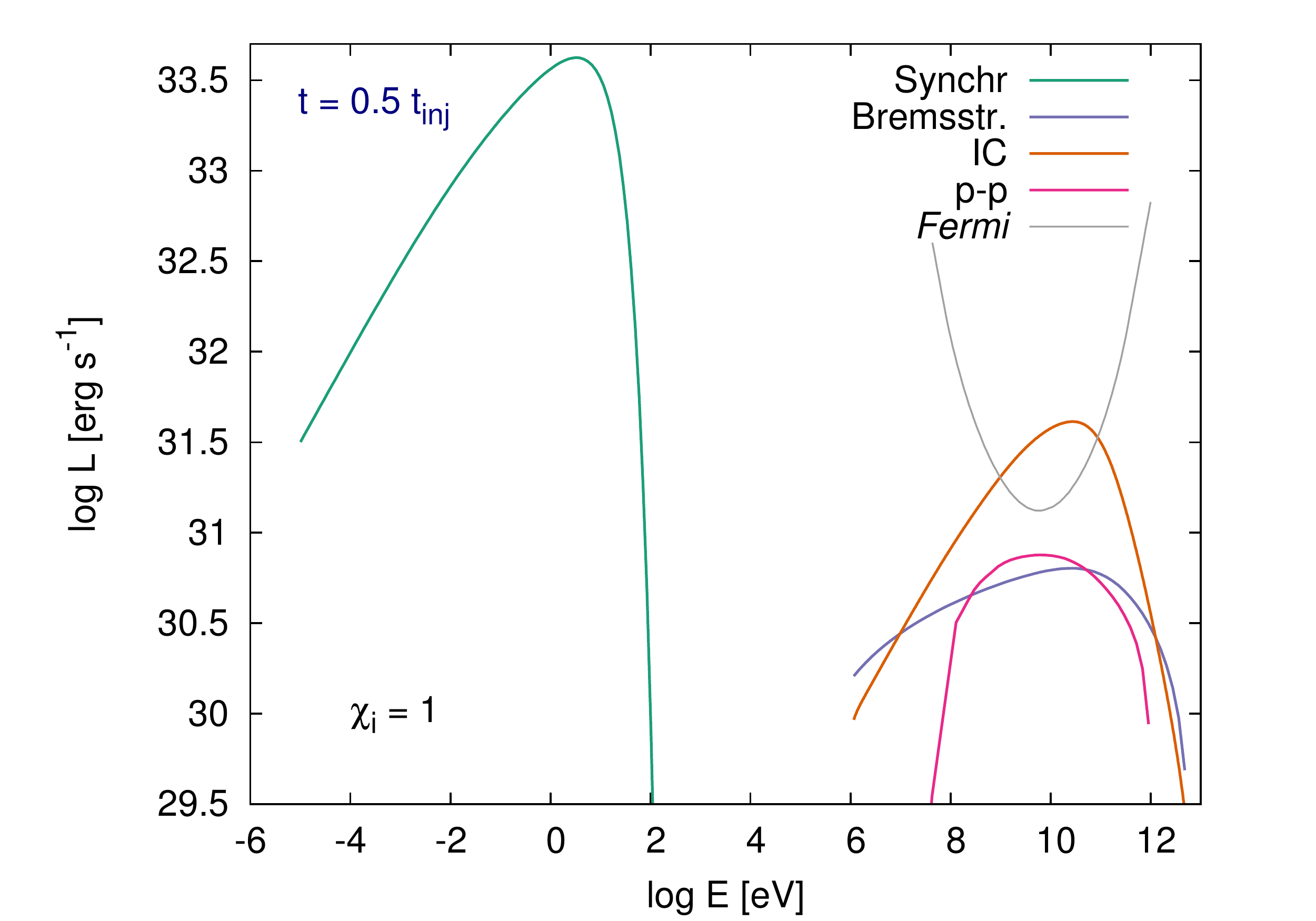}}
\resizebox{.67\columnwidth}{!}{\includegraphics[scale=.8, trim=0cm 0cm 0cm 2.2cm, clip=true,angle=270]{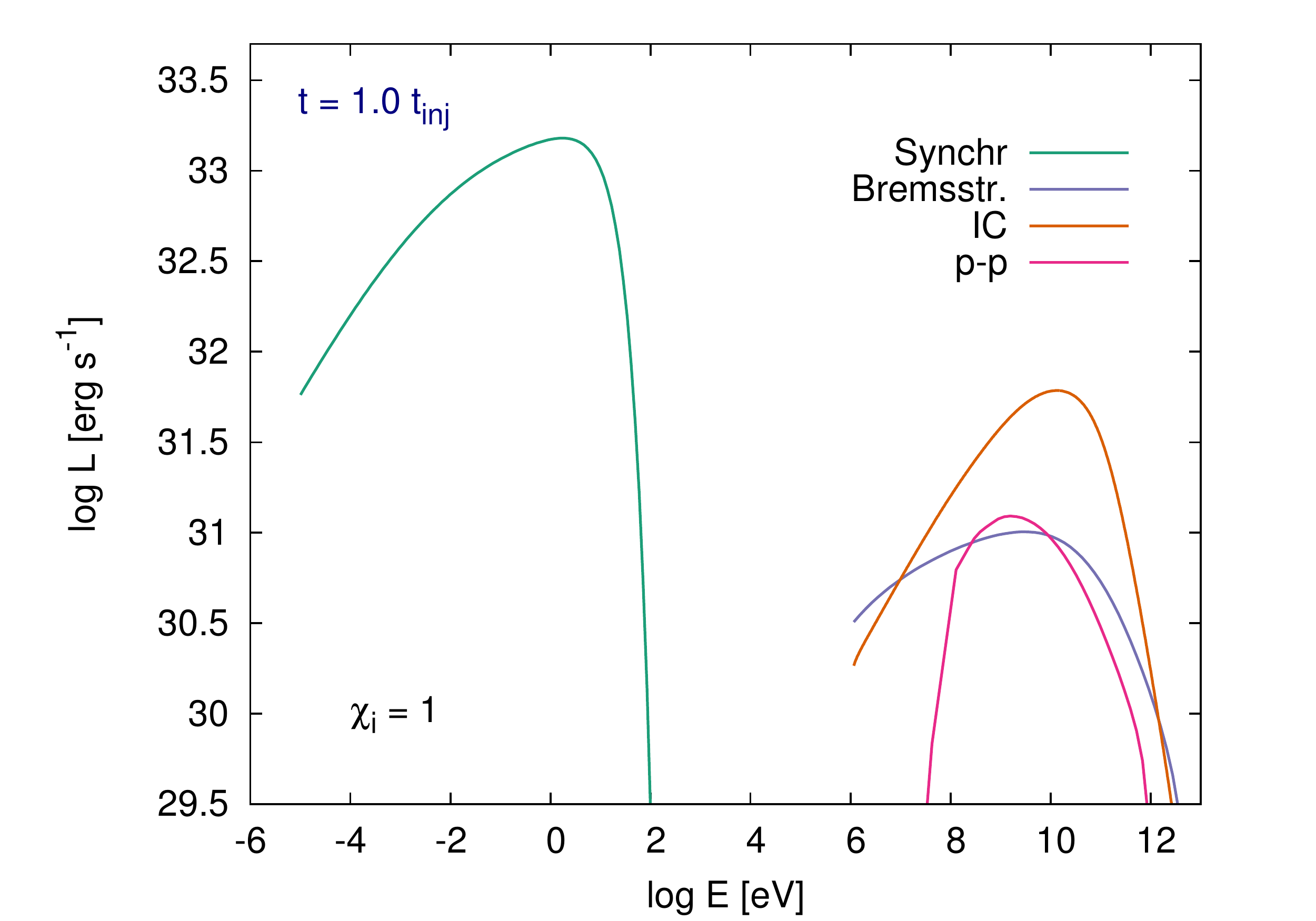}}\\
\resizebox{.67\columnwidth}{!}{\includegraphics[scale=.8,trim=0cm 0cm 0cm 2.2cm, clip=true,angle=270]{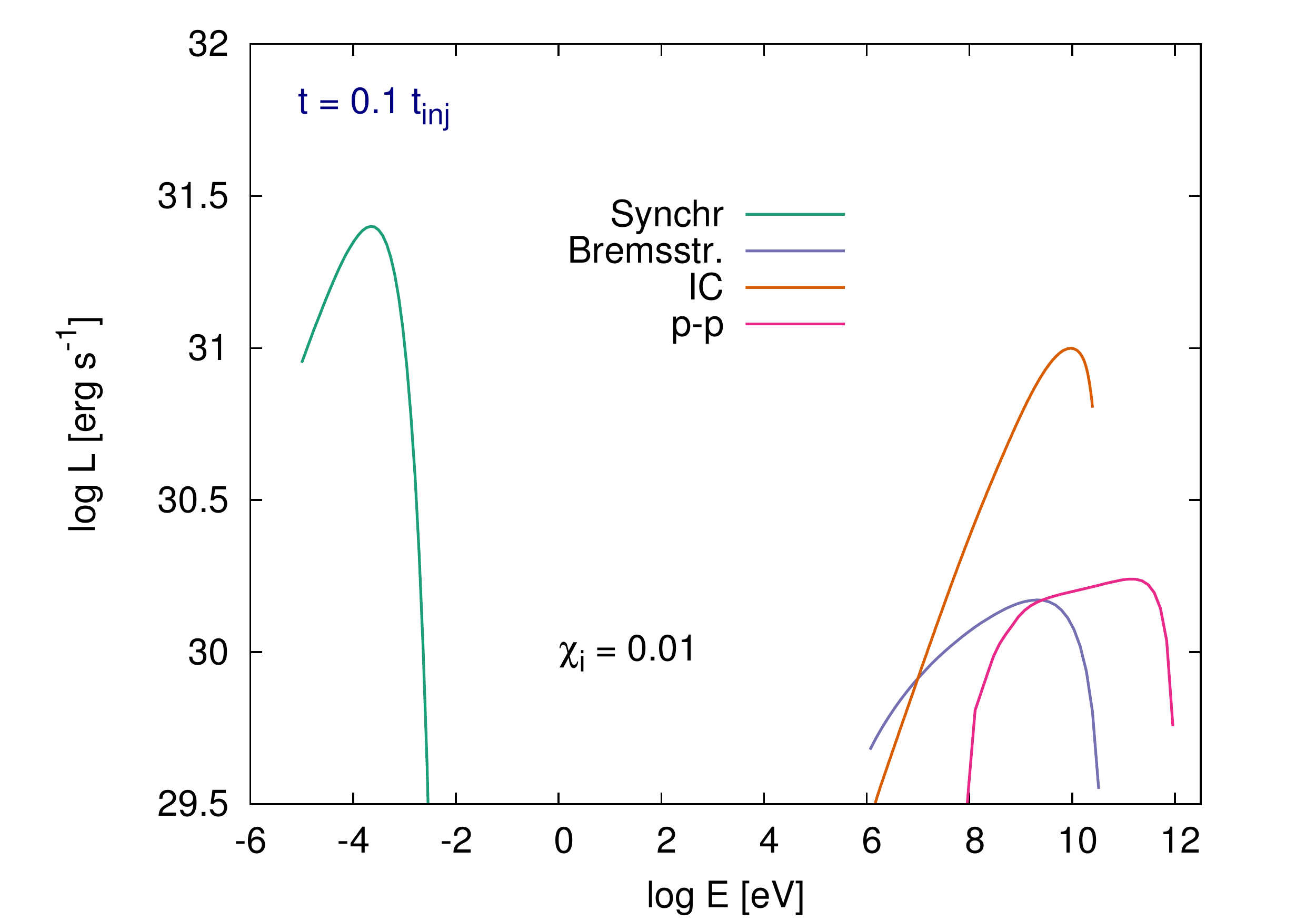}}
\resizebox{.67\columnwidth}{!}{\includegraphics[scale=.8, trim=0cm 0cm 0cm 2.2cm, clip=true,angle=270]{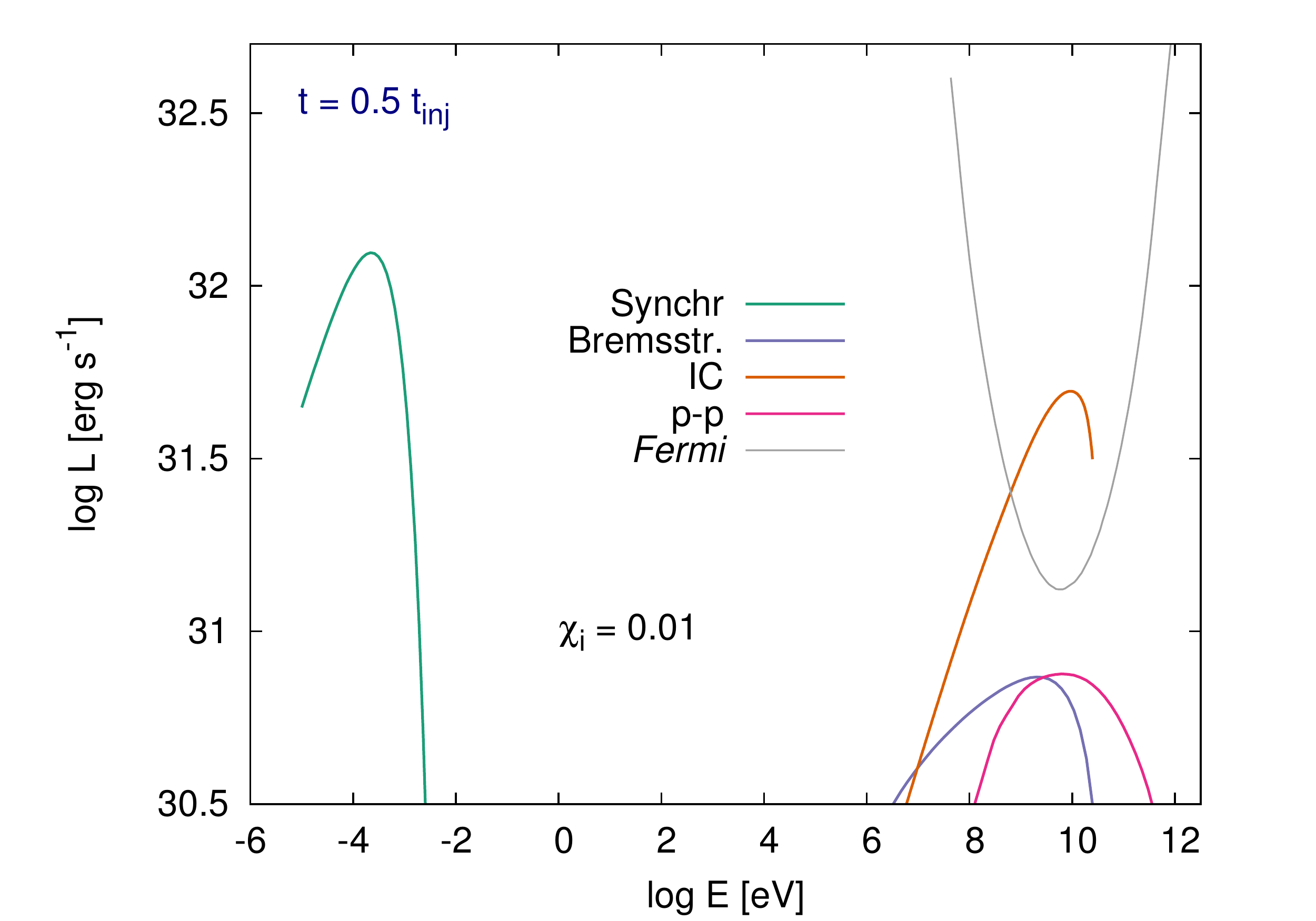}}
\resizebox{.67\columnwidth}{!}{\includegraphics[scale=.8, trim=0cm 0cm 0cm 2.2cm, clip=true,angle=270]{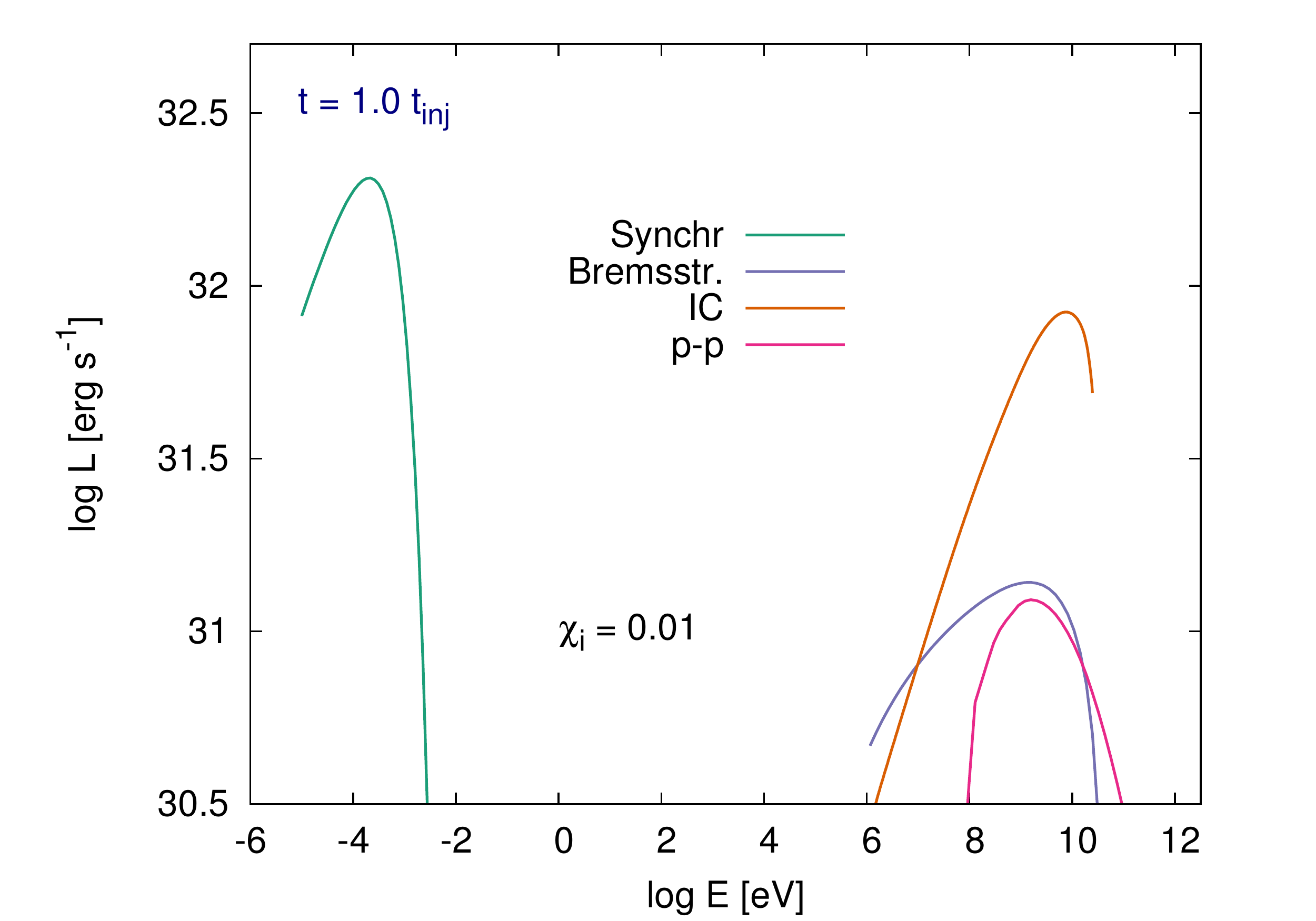}}
\caption{ Spectral energy distribution for three different  injection times for Model \emph{D} in the case of a fully ionized disk, $\chi_{\rm i} \sim 1$, (top) and an almost neutral disk, $\chi_{\rm i} \sim 0.01$, (bottom). The left panel shows the non-thermal SED for $t = 0.1\,t_{\rm inj}$, the middle panel shows the SED at  $t = 0.5\,t_{\rm inj}$ and the right panel corresponds to the final integration time $t = 1.0\,t_{\rm inj}$.}
\label{fig:sed-caso1}
\end{center}
\end{figure*}

The corresponding non-thermal SEDs for the cases $\chi_{\rm i} \sim 1$ and $\chi_{\rm i} \sim 0.01$ are shown in Fig.\,\ref{fig:sed-caso1}, for 3 different times during the integration period. We only calculate the IC  emission due to interactions with the cosmic background, because this is by far the dominant contribution (see Fig.\ref{fig:losses}). The emission reaches its maximum when the injector is located in the middle of the integration region. The contribution from protons, as expected, is identical in both cases. The main contributions come from the electrons, with a higher luminosity at radio wavelengths due to the synchrotron radiation that is of the order of $10^{33}$\,erg\,s$^{-1}$ for $\chi_{\rm i} \sim 1$ and $10^{32}$\,erg\,s$^{-1}$ for $\chi_{\rm i} \sim 0.01$. This difference in the luminosity is because in the neutral case there is a lack of energetic electrons. Also, the synchrotron emission reaches smaller energies in the neutral disk case, as expected from an electron population with lower maximum energy. The gamma-ray emission is dominated by IC, reaching a maximum luminosity of $\sim$ $5 \times 10^{31}$\,erg\,s$^{-1}$ at soft gamma rays in both cases, but in the case with  $\chi_{\rm i} \sim 0.01$ this maximum is slightly shifted to lower energies. The contributions from relativistic Bremsstrahlung and from $p-p$ collisions are negligible.

In the middle panel of Fig.\,\ref{fig:sed-caso1}, in grey, we plot the {\it Fermi} sensitivity curve  for a source at $d \sim 1 $\,kpc. The IC luminosity lies above the curve, even in the case of a single collision. Hence the gamma radiation from an event of these characteristics might be detectable at energies around $\sim$ 10\,GeV  for both $\chi_{\rm i} \sim 1$ and $\chi_{\rm i} \sim 0.01$. Some unidentified  {\it Fermi} sources might correspond to this kind of objects. In the case of multiple events, as discussed in the previous scenario,  the luminosity can increase one order of magnitude, making the event detectable on a wider energy range. Also, the collective emission might be detectable, at least around the energies of maximum {\it Fermi} sensitivity, at greater distances,  say up to $d \sim 5 $\,kpc.

The protons lose only a small fraction of their energy by $p-p$ in the cloud, they diffuse into the surroundings, adding their energy to the local cosmic-ray sea. In what follows we briefly discuss the implications of this. 

\subsection{Discussion}

In order to compare locally the injected protons with those from the cosmic-ray background we calculate this latter contribution as explained in Sect.\,\ref{sec:part-disk} above; the results are shown over the maps in Figure\,\ref{fig:Np-map-disk}. The white dashed curves indicate the region  where the local proton flux is equal to the background CR protons, hence in the region inside the curve the  protons accelerated at the shocked disk exceed the background. This region, around the injection axis,  is greater for later times, and it extends from the axis  $\sim$ 10\,pc for $E = 1$\,GeV up to  $\sim$ 33\,pc in the case of higher energies. 

The locally injected protons dominate over the background on a  considerable region, especially  at high energies. This is expected because of the softer distribution of cosmic rays compared with  the local particles. As a result the shock propagation through the disk enhances the local number of energetic protons. This effect can be even stronger when considering collective effects, i.e. a bigger HVC fragmenting in smaller pieces of the size studied here (see Sect.\,\ref{sec:discu}). Such a case is illustrated by the golden (outer) dashed curves in Figure\,\ref{fig:Np-map-disk}, where we consider that the injected number of  protons is one order of magnitude higher. The region where the local flux dominates over the background extends beyond  66\,pc  for protons of $E = 1$\,TeV. 

The rate of mass injected in the Galaxy by HVCs impacts is 0.5\,M$_{\odot}$\,Yr$^{-1}$ \citep{richter12}. The total power released, if all the clouds have the same velocity $V_{\rm c}$, is: 

\begin{equation}
  P_{\rm HVCs} \sim 2 \times 10^{40} \left(\frac{V_{\rm c}}{250\,{\rm km}\,{\rm s}^{-1}}\right)^2\, {\rm erg}\,{\rm s}^{-1}.
\end{equation}

\noindent Assuming that half of this power is transfer in the collisions to disk-propagating  shocks we can estimate the power in relativistic particles. If the shocks accelerate particles with a $10\%$ efficiency then for an average cloud velocity of $V_{\rm c} \sim 250$\,km\,s$^{-1}$, the power in relativistic particles is $P^{\rm CR}_{\rm HVCs} \sim 10^{39}\, {\rm erg}\,{\rm s}^{-1}$. This power is non -negligible considering that the power in the Galaxy from CRs is $\sim$ $5\times 10^{40}{\rm erg}\,{\rm s}^{-1}$. We conclude that up to 10\% of the CRs in the Galaxy can be produced by impacts of HVCs.

\section{Summary and conclusions}
\label{sec:concl}

In this work we investigated the non-thermal effects of the collision of a HVC with the Galactic disk. We analyzed the properties of the shocks produced in the interaction, concluding that under some general conditions DSA can operate efficiently. We considered two different scenarios for computing the non-thermal emission: a reverse shock propagating through the cloud and a forward shock propagating through the Galactic disk. 

We solved the transport equation for electrons and protons in a spherical cloud, for two different sets of parameters. We found that significant non-thermal radio emission occurs, with maximum luminosities around $\sim$ $10^{32}$\,erg\,s$^{-1}$; we also found a moderate soft gamma component,  with maximum power of the order of $10^{29}$\,erg\,s$^{-1}$. In the case of multiple impacts the corresponding luminosities can be an order of magnitude higher.

For the case of an adiabatic shock propagating through the disk we solved the transport equation of particles in cylindrical coordinates. The leptonic contributions dominate the SEDs, with synchrotron radiation being at the level of  $\sim$ $10^{33}$\,erg\,s$^{-1}$
and gamma-ray emission, from IC up-scattering of the cosmic background, at $\sim$ $10^{31}$\,erg\,s$^{-1}$. We also compared the number of protons that are locally accelerated with the Galactic cosmic-ray population, finding that the local component of protons (especially those of the highest energies) dominates over the background on a substantial region. 

The impact of HVCs with the Galactic disk releases a great amount of energy into the ISM; we showed that under some reasonable conditions a fraction of that energy can be converted into non-thermal radiation and energetic particles. Furthermore, under some circumstances the non-thermal emission might be detectable and  the power in energetic particles might form a non-negligible contribution of the global population of Galactic CRs.


\section*{Acknowledgements}
The authors would like to thank Dr. Reinaldo Santos-Lima for his help on numerical affairs. M.~V.~d.V. acknowledges partial support from the Alexander von Humboldt Foundation. G.~E.~R acknowledges support from the Argentine Agency CONICET (PIP 2014-00338) and the Spanish Ministerio de Econom\'{i}a y Competitividad (MINECO/FEDER, UE) under grants AYA2013-47447-C3-1-P and AYA2016-76012-C3-1-P.




\bibliographystyle{mnras}




\appendix
\section{Transport of energetic particles}\label{appendix}

In order to solve the transport of relativistic particles and to calculate the subsequent non-thermal emission we have used a modular code presented in \citet[][]{delvalle15}. The transport equations, Eqs.\,(\ref{eq:transcloud}) and (\ref{eq:transdisc}),  were solved in spherical and cylindrical coordinates, respectively, allowing a reduction of the dimensionality of the problem given the existing symmetries.

The transport equations for electrons and protons were evolved simultaneously using the finite-volumes method. We adopted a discrete grid $(E, R, \theta)$ in spherical coordinates and $(E, r, z)$
in cylindrical coordinates; where   $(R, \theta) / (r,z) $ are the usual spatial coordinates in the correspondent system and  $E$ is the energy of the particles. The energy grid was logarithmically spaced, whereas the spatial grids were sampled uniformly.

Initially, at time $t=0$,  we considered no particles inside the domain.  We imposed that there were no particles outside the energy bounds. These limits did not influence the system evolution, 
because the upper limit was above the maximum energy of the injected particles, 
at the same time that the advection in the energy space (the term of energy losses) is always towards smaller energies.  The lower bound was physically fixed by the particles rest mass. 

For the spatial boundary conditions we considered zero particles outside the domain, i.e. for $R > r_{\rm c}$ in the spherical case, and for $r > r_{\rm d}$ and $z > r_{\rm d}$ in cylindrical coordinates. Because of the azimuthal symmetry in the case of spherical coordinates at $\theta = 0$ and $\theta=\pi$, we imposed there outflow boundary conditions. In the case of cylindrical  coordinates, at the inner boundary condition for $r$  we adopted axial symmetry. 

The numerical integration was performed through the operator splitting method. Each time-step integration evolved the particle density distribution on the grid through three sub-steps: first we integrated the losses, then the spatial diffusion and finally we added the source term.

The loss term is an advection in energy space, therefore to solve it we employed the finite-volume 
formulation with an upwind scheme of second order. In order to calculate the fluxes at the interface of the cells we used the Piecewise Linear Method (PLM) with the monotonic Central limiter, which is second order accurate.
The intermediate solution was then obtained through the explicit Euler method. 

For integrating the spatial diffusion part of the transport equation we applied the semi-implicit Cranck-Nicolson method, with the gradients calculated at the cell interfaces, using central differences. This scheme is, therefore, second order accurate. 

As a last step the contributions from the injection term were added, using the Euler explicit method.

The time-steps were chosen in accordance with the Courant-Friedrichs-Lewy (CFL) stability criterion for the minimum time step of the advection and diffusion equations. Additionally, we imposed the condition that the time step must be smaller than the time the injector takes to cross one cell. 




\bsp	
\label{lastpage}
\end{document}